 \documentclass[twocolumn,trackchanges]{aastex631}

\usepackage{float}
\usepackage{chemformula}

\shorttitle{NIR variability of WISE 1049AB}
\shortauthors{Oliveros-Gomez et al.}
\graphicspath{{./}{figures/}}
\usepackage{booktabs}
\usepackage{rotating}
\usepackage{verbatim} 
\usepackage[normalem]{ulem}
\usepackage{booktabs, longtable}
\usepackage{threeparttable} 
\usepackage{nicefrac}
\usepackage[version=4]{mhchem}
\usepackage{xcolor}

\begin{document}

\title{The JWST weather report from the nearest brown dwarfs III: Heterogeneous clouds and Thermochemical instabilities as possible drivers of WISE 1049AB’s spectroscopic variability}

\author[0000-0001-5254-6740]{Natalia Oliveros-Gomez}
\affiliation{William H. Miller III Department of Physics and Astronomy, Johns Hopkins University, Baltimore, MD 21218, USA}

\author[0000-0003-0192-6887]{Elena Manjavacas}
\affiliation{AURA for the European Space Agency (ESA), ESA Office, Space Telescope Science Institute, 3700 San Martin Drive, Baltimore, MD, 21218 USA}
\affiliation{William H. Miller III Department of Physics and Astronomy, Johns Hopkins University, Baltimore, MD 21218, USA}

\author[0000-0001-7356-6652]{Theodora Karalidi}
\affiliation{Department of Physics, University of Central Florida, 4000 Central Florida Blvd., Orlando, FL 32816, USA}

\author[0000-0001-6059-9471]{Myrla Phillippe}
\affiliation{Department of Physics, University of Central Florida, 4000 Central Florida Blvd., Orlando, FL 32816, USA}

\author[0000-0003-0225-1201]{Beatriz Campos Estrada}
\affiliation{Max-Planck-Institut fur Astronomie, Konigstuhl 17, D-69117 Heidelberg, Germany}

\author[0000-0003-4614-7035]{Beth Biller}
\affiliation{Institute for Astronomy, University of Edinburgh, Royal Observatory, Edinburgh EH9 3HJ, UK}
\affiliation{Centre for Exoplanet Science, University of Edinburgh, Edinburgh, UK}

\author[0000-0003-0489-1528]{Johanna M. Vos}
\affiliation{School of Physics, Trinity College Dublin, The University of Dublin, Dublin 2, Ireland}
\affiliation{Department of Astrophysics, American Museum of Natural History, Central Park West at 79th Street, NY 10024, USA} 

\author[0000-0001-6251-0573]{Jacqueline Faherty}
\affiliation{Department of Astrophysics, American Museum of Natural History, Central Park West at 79th Street, NY 10024, USA}

\author[0009-0005-9339-2369]{Xueqing Chen}
\affiliation{Institute for Astronomy, University of Edinburgh, Royal Observatory, Edinburgh EH9 3HJ, UK}
\affiliation{Centre for Exoplanet Science, University of Edinburgh, Edinburgh, UK} 

\author[0000-0001-9823-1445]{Trent J. Dupuy}
\affiliation{Institute for Astronomy, University of Edinburgh, Royal Observatory, Edinburgh EH9 3HJ, UK}
\affiliation{Centre for Exoplanet Science, University of Edinburgh, Edinburgh, UK}

\author[0000-0002-1493-3000X]{Thomas Henning}
\affiliation{Max-Planck-Institut fur Astronomie, Konigstuhl 17, D-69117 Heidelberg, Germany} 

\author[0000-0003-2015-5029]{Allison M. McCarthy}
\affiliation{Department of Astronomy \& The Institute for Astrophysical Research, Boston University, 725 Commonwealth Ave., Boston, MA 02215, USA} 
\affiliation{School of Physics, Trinity College Dublin, The University of Dublin, Dublin 2, Ireland}

\author[0000-0002-0638-8822]{Philip S. Muirhead}
\affiliation{Department of Astronomy \& The Institute for Astrophysical Research, Boston University, 725 Commonwealth Ave., Boston, MA 02215, USA} 

\author[0000-0002-3052-7116]{Elspeth K. H. Lee}
\affiliation{Center for Space and Habitability, University of Bern, Gesellschaftsstrasse 6, CH-3012 Bern, Switzerland} 

\author[0000-0001-6172-3403]{Pascal Tremblin}
\affiliation{Universit\'{e} Paris-Saclay, UVSQ, CNRS, CEA, Maison de la Simulation, 91191 Gif-sur-Yvette, France} 

\author[0009-0002-1115-7700]{Jasmine Ramirez}
\affiliation{The Graduate Center, City University of New York, New York, NY 10016, USA}
\affiliation{Department of Astrophysics, American Museum of Natural History, Central Park West at 79th Street, New York, NY 10034, USA}

\author[0000-0002-2011-4924]{Genaro Suarez}
\affiliation{Department of Astrophysics, American Museum of Natural History, Central Park West at 79th Street, New York, NY 10034, USA}

\author[0000-0002-9962-132X]{Ben J. Sutlieff}
\affiliation{Institute for Astronomy, University of Edinburgh, Royal Observatory, Edinburgh EH9 3HJ, UK}
\affiliation{Centre for Exoplanet Science, University of Edinburgh, Edinburgh, UK}

\author[0000-0003-2278-6932]{Xianyu Tan}
\affiliation{Tsung-Dao Lee Institute \& School of Physics and Astronomy, Shanghai Jiao Tong University, 1 Lisuo Road, Shanghai, 201210, People’s Republic of China} 

\author[0000-0001-7866-8738]{Nicolas Crouzet}
\affiliation{Kapteyn Astronomical Institute, Rijksuniversiteit Groningen, Postbus 800, 9700 AV Groningen, The Netherlands}

\begin{abstract}

We present a new analysis of the spectroscopic variability of WISE~J104915.57$-$531906.1AB (WISE~1049AB, L7.5+T0.5), observed using the NIRSpec instrument onboard the James Webb Space Telescope (GO 2965 - PI: Biller).
We explored the variability of the dominant molecular bands present in their 0.6--5.3~$\mu$m spectra (H$_2$O, CH$_4$, CO), finding that the B component exhibits a higher maximum deviation than the A component in all the wavelength ranges tested. The light curves reveal wavelength-(atmospheric depth) and possibly chemistry-dependent variability. In particular, for the A component, the variability in the light curves at the wavelengths traced by the CH$_4$ and CO molecular absorption features is higher than that of H$_2$O, even when both trace similar pressure levels. We concluded that clouds alone are unlikely to explain the increased variability of CO and CH$_4$ with respect to H$_2$O, suggesting that an additional physical mechanism is needed to explain the observed variability. This mechanism is probably due to thermochemical instabilities. 
Finally, we provide a visual representation of the 3D atmospheric map reconstructed for both components using the molecular band contributions at different pressure levels and the fit of planetary-scale waves.   
\end{abstract}



\section{Introduction} \label{sec:intro}

Brown dwarfs of all spectral types show some degree of photometric or spectroscopic variability, with periods that range from a few hours to more than 24~hr (e.g., \citealt{buenzli2014brown}, \citealt{radigan2014independent}, \citealt{metchev2015weather}, \citealt{cushing2016first}, \citealt{leggett2016observed}, \citealt{Vos2022}). However, the maximum deviation of the variability, i.e., the percent of relative variability between the highest peak and deepest trough observed in a light curve, of L/T transition brown dwarfs is usually higher than the rest of the spectral types of brown dwarfs \citep{radigan2014strong}. This increase in maximum deviation might be caused by the presence of patchiness in silicate clouds that have sedimented, which are near-ubiquitous in L dwarfs, from the atmosphere over the transition between L and T spectral types (\citealt{burgasser2006method}, \citealt{looper2008discovery}). Species like MgSiO$_3$, Mg$_2$SiO$_4$, and $\mathrm{Na_{2}S}$ condense in the atmospheres of brown dwarfs at different altitudes, depending on their pressure-temperature profiles, generating various types of clouds (\citealt{ackerman2001precipitating}, \citealt{Tan_Showman2021}, \citealt{vos2023patchy}). As these patchy clouds rotate in and out of view, we detect wavelength-dependent brightness changes. Weather patterns of heterogeneous cloud coverage in brown dwarf atmospheres have been considered the most likely mechanism explaining photometric variability (e.g., \citealt{apai2013hst}, \citealt{miles2017weather}). 

Other mechanisms have been proposed to explain the observed atmospheric variability in brown dwarfs, namely disequilibrium chemistry. At the L/T transition, the equilibrium reaction of carbon that takes place is: 

\begin{equation}
\ce{CO + 3H2 <->CH4 +H2O}.
\label{eq:reaction}
\end{equation}

At higher temperatures, the left side of the reaction in equation \ref{eq:reaction} is favored (\citealt{Lodders2002atmospheric}, \citealt{Zahnle2014methane}); thus, there is an overabundance of CO, which implies an underabundance of $\mathrm{H_{2}O}$ and CH$_4$. At lower temperatures, the right-hand side of the reaction is favored, leading to higher abundances of $\mathrm{CH_{4}}$ and $\mathrm{H_{2}O}$.  Disequilibrium chemistry occurs when these chemical reactions do not reach equilibrium in the atmosphere. This is due to atmospheric convection and turbulence that bring up gas from deeper and hotter atmospheric layers, where CO is stable. This prevents CO from being converted to CH$_4$ at the expected equilibrium levels.  Disequilibrium chemistry causes brown dwarfs to show an excess of CO and a deficit of CH$_4$ in their spectra compared to what equilibrium chemistry predicts (e.g., \citealt{miles2020observations}). Variability can thus also potentially arise from chemical disequilibrium, where hotter/colder regions in which chemical abundances differ come in and out of view as the brown dwarf rotates.

Recent 3D Global Circulation Models (GCMs) which consider both chemical and cloud feedback in the atmospheric properties, show that temperature changes and dynamical motions induced by cloud opacity can trigger convection, which in turn drives chemical disequilibrium, and therefore increased atmospheric variability \citep{lee2024dynamically}.

However, \citet{tremblin2016cloudless, tremblin2020rotational} proposed that carbon thermochemical instabilities triggered by CO/$\ce{CH4}$ radiative convection alone can explain variability without the need for the presence of clouds.

At the L/T transition, this is particularly important, as the rate of the CO/CH$_4$ chemical reaction may affect the observed variability, producing ``patches" of disequilibrium chemistry caused by temperature fluctuations in some parts of the atmosphere. These mechanisms theoretically play an essential role in the atmospheres of brown dwarfs and have been verified by models that include thermochemical instabilities and provide best fits of brown dwarf spectra in comparison to the models that include just clouds \citep{tremblin2016cloudless}. However, confirming this mechanism with variability observations has been challenging, because they do not have the pertinent signal-to-noise (SNR) and wavelength coverage to analyze properly.

With the launch of the James Webb Space Telescope (JWST) came a new generation of spectrographs that allow us to obtain brown dwarf spectra covering the near-infrared wavelengths where the CO and CH$_4$ bands.
Thanks to the capabilities of JWST/NIRSpec, it is now possible to measure the spectroscopic variability in key molecular absorption features present in the atmospheres of brown dwarfs, such as CH$_4$, CO, and H$_2$O in the 0.6--5.3\,$\mu$m range. We can then compare the maximum deviation of the different molecular absorption features, allowing us to study the role of different variability mechanisms, such as inhomogeneous distribution of clouds and thermochemical instabilities in the atmosphere.

In addition to understanding the different variability mechanisms, time-resolved spectroscopy over a wide wavelength range of brown dwarf atmospheres allows us to provide a 3D characterization of their atmospheres and weather patterns. Using photometric or spectroscopic variability with a sufficiently long monitoring baseline, covering at least a few rotations of the object, it is possible to make indicative maps of top-of-atmosphere structure (TOA) with codes such as \textit{Stratos} \citep{apai2013hst}, \texttt{Aeolus} \citep{karalidi2015aeolus}, that are physically motivated, and \textit{Starry} \citep{Luger2019}, that follow data-driven/agnostic methods. These models allow us to derive 2D or 3D brightness distributions that are consistent with the data and may provide exploratory maps of the atmospheres, in terms of structures, such as bands and spots of clouds in the TOA.

In this paper, we aim to understand the different physical mechanisms that explain the spectroscopic variability measured in the 0.6--5.3\,$\mu$m range in the WISE J104915.57$-$531906.1AB system (henceforth WISE~1049AB). The paper is structured as follows: Section~\ref{sec:objects} summarizes the previous studies of WISE~1049AB. Section~\ref{sec:observations} provides the details of the NIRSpec/PRISM BOTS onboard JWST WISE~1049AB observations. Section~\ref{sec:data_analysis} explains the methodology followed to study the chemistry contribution in the atmospheres, and discusses the qualitative maps of both objects. In Section~\ref{sec:depth_dependent_variability}, we discuss the depth-dependent variability for different species, the different mechanisms that might explain the variability measured, and we provide a visual illustration of the atmospheric structures that might be present in WISE~1049AB. In Section~\ref{sec:discussion}, we compare and discuss our results with previous studies. Finally, in Section~\ref{Sec:conclusions}, we summarize our conclusions.

\section{WISE J104915.57-531906.1AB}\label{sec:objects}

WISE~1049AB are the two brightest ($J~=~10.73$\,mag, \citealt{luhman2013discovery}) and closest ($1.9960\,{\rm pc} \pm 50$\,AU, \citealt{bedin2024hst}) brown dwarfs to Earth. WISE~1049AB is a spatially resolved binary system, with a projected separation of approximately 3.5\,AU with an orbital period of about 30\,yr \citep{garcia2017individual}. Both objects are in the L/T transition, with spectral types of L7.5 for A and T0.5 for B \citep{burgasser2013resolved}. Thanks to its proximity and brightness, WISE~1049AB has been photometrically and spectroscopically monitored for over a decade, and their orbits and dynamical masses are well-constrained ($35.4\pm0.3\,M\mathrm{_{Jup}}$ for the A component and $29.4 \pm0.2\,M\mathrm{_{Jup}}$, for the B component \citealt{bedin2024hst}). 

\cite{biller2024jwst} calculated the effective temperatures ($T_{\mathrm{eff}}$) and gravities ($\log{g}$) for both components of the system from bolometric luminosities (derived from this dataset and the accompanying MIRI dataset of the same program) and evolutionary model radii. They obtained $T_{\mathrm{eff}}$ between 1150~K and 1300~K and $\log{g}$ between 4.7\,dex and 5\,dex. The binary system is part of the Oceanus moving group, with an estimated age of 500~Myr \citep{gagne2023moving}.

The first variability study of this system by \cite{gillon2013fast} was not able to separate the contributions of two objects, but they found a quasiperiodic variability with a period of 4.87 hr, and an amplitude of $\sim$11\% peak-to-peak, which they attributed to the existence of heterogeneous clouds in the atmospheres. \cite{biller2013weather} was able to separate for the first time the contributions of each component, finding tentative variability for the A component, and a higher variability for the B component ($\sim$10-13 \% in the $H$- and $J$- bands).

Follow-up studies by \cite{burgasser2014monitoring} found a peak-to-peak amplitude of 4.5\% using the TRAPPIST broadband $I$+$z$ filter over the combined light curve of both objects and a period of 5.05\,hr. In addition, using low-resolution near-infrared spectroscopic monitoring observations, they found 13.5\% of variability at 1.25\,$\mu$m. They inferred a brightness temperature of 1560\,K, using a simple two-spot brightness temperature model for WISE~1049B, with variations of 200-–400\,K between hot and cold regions. 

\cite{buenzli2015cloud} characterized the time-resolved spectroscopy of the system using the Hubble Space Telescope (HST) Wide Field Camera 3 (WFC3) over 1.10--1.66\,$\mu$m for a duration of 6.5\,hr. They measured a variability amplitude of 0.4\% for WISE~1049A and a peak-to-valley variability amplitude of 7\%- 11\% for WISE~1049B. \cite{buenzli2015cloud} also reproduced its spectrum using a single layer of clouds \citep{marley2012masses} with an intermediate thickness of the cloud assuming chemistry equilibrium ($f_\mathrm{sed}$ = 2, $T_\mathrm{eff} = 1200$\,K, $\log g = 4.5$\,dex, $K_\mathrm{zz} = 10^4$, $R = 0.95~\mathrm{R_{Jup}}$). 

\citet{crossfield2014global} produced a two-dimensional map of the WISE~1049B TOA via Doppler imaging, generated via high-resolution spectroscopy observations that enable different top-of-atmosphere features to be distinguished at various latitudes and longitudes on the brown dwarf, and allowed the identification of large-scale bright and dark features indicative of patchy clouds. They constrained the inclination of WISE1049B to $i = 30^\circ$. In addition, \citet{chen2024global} obtained the first global Doppler map of WISE 1049A and also a map of B object, using Gemini IGRINS spectroscopy (1.5--2.5\,$\mu$m), for the \textit{H} and \textit{K} bands, where they used cloudless/cloudy models, and chemical equilibrium to investigate the atmospheric structures that give rise to the variability. \citet{chen2024global} suggests that atmospheric cold spots cannot be due to chemical hot spots alone but must involve clouds.

On the other hand, \cite{karalidi2016maps} created a map of WISE~1049AB using  HST/WFC3 light curves and the \texttt{Aeolus} code \citep{karalidi2015aeolus}. \cite{karalidi2016maps} found that three or four patchy clouds in the TOAs of WISE~1049A and B were required to reproduce the variability data, with 19\%--32\% and 21\%--38.5\% surface coverage, respectively. The \texttt{Aeolus} code retrieved an inclination of $56^\circ \pm 5^\circ$ for WISE~1049A, and $26^\circ \pm 8^\circ$ for WISE~1049B (considering $0^\circ$ as equator-on and $90^\circ$ as pole-on), consistent with the inclination calculated by \citet{crossfield2014global}.

\cite{Apai2021} used the Transiting Exoplanet Survey Satellite (TESS) photometric data to obtain a light curve of WISE~1049AB, covering about 100 rotations of the system. Their periodogram analysis showed that the B component's rotational period of 5.28\,hr dominates the light curve in the TESS band. They argued that a secondary 6.94\,hr period present in the light curve is likely the rotation period of WISE~1049A. Additionally, an analysis of equatorial rotational velocities suggests a viewing angle close to the equator for both brown dwarfs. According to \cite{Apai2021}, WISE~1049A is viewed from an angle within $28^\circ$ of its equatorial plane (i.e., $i > 62^\circ$), and WISE~1049B is viewed almost exactly equatorially (within a few degrees, i.e., $i \sim 90^\circ$), consistent with the inclination calculated in previous studies (\citealt{crossfield2014global} and \citealt{karalidi2016maps}).  \cite{fuda2024latitude} extended this study to include 1200\,hr of photometric observations with TESS and found variability of $\pm 5\% $ on a time scale of several to hundreds of hours. They also showed that the short-period rotational modulation around 2.5--5\,hr dominates the variability below 10 hours. They used composite sine-wave models to explain the variability of light curve durations up to 100 hours, where each function modulates cloud thickness, producing rotational modulation in the rotating atmospheres.

\section{Observations}\label{sec:observations}

Our data, presented in \citet{biller2024jwst}, were obtained with JWST/NIRSpec BOTS (Bright Object Time-Series) spectroscopic mode. We obtained time-resolved spectroscopy using the low-resolution PRISM/CLEAR mode. The spectra were obtained on UT 2023 July 8 21:37:21 to UT 2023 July 9 05:12:53, as part of JWST Cycle~2 (GO-2965; PI: Biller). This observing mode provides simultaneous wavelength coverage from 0.6 to 5.3\,$\mu$m at a resolution ranging from R~=~30 to 300.

WISE 1049AB was observed continuously during $\sim$8~hr,  covering at least one complete period for each object (WISE~1049A has a rotational period $P\approx7$ hr and WISE~1049B has a rotational period $P\approx5$ hr, \citealt{Apai2021}).  Since WISE~1049AB  are high-brightness brown dwarfs, we observed the target using two groups and 57,100 integrations to avoid saturation, obtaining $\sim$57\,000 spectra for each object. WISE~1049AB was observed with a V3 PA (observatory V3 reference axis position angle) of 130.6$^\circ$, corresponding to a NIRSpec aperture position angle (APA) of 269.8$^\circ$. When the system was monitored, WISE~1049B was at a separation of $0.65''$ and a position angle of 113$^\circ$ relative to WISE~1049A, as measured by NIRSpec acquisition images. Thus, the traces of both objects in NIRSpec are well resolved ($>5$ pixels separation between traces). The details of the data reduction are explained in \cite{biller2024jwst}. We used the same reduced data presented there.

\vspace{15mm}
\section{Data Analysis}\label{sec:data_analysis}

\subsection{WISE 1049AB Spectra and Spectral Ratios}

\begin{figure*}
    \centering
    \includegraphics[width=1\linewidth]{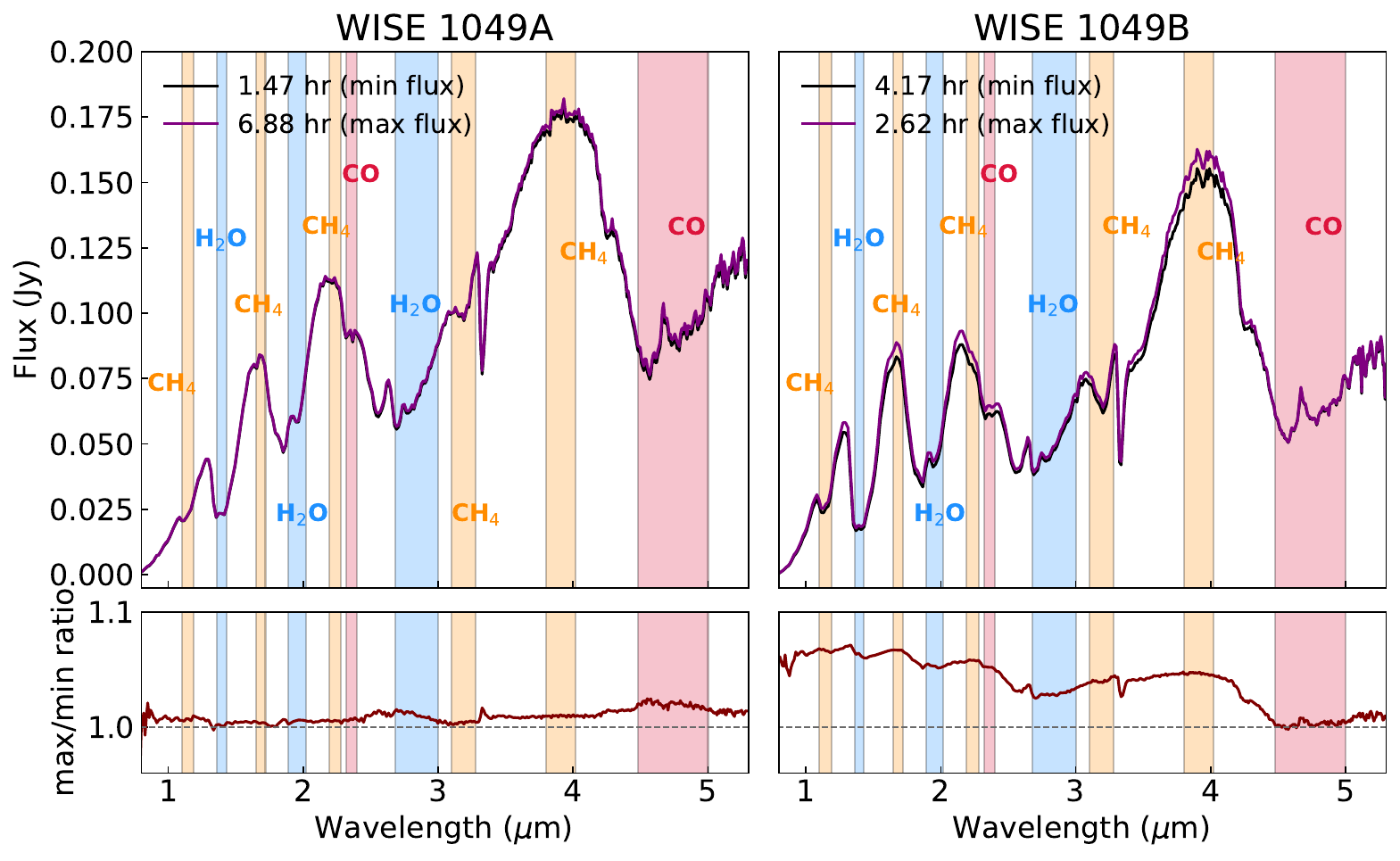}
    \caption{\textit{Upper panel:} Median of the 21 maximum spectra (purple line) and median of the 21 minimum spectra (black line) of WISE 1049A in left panel and WISE 1049B in right panel, taken from ``white light curve" (0.6 -- 5.3 $\mu$m) 7 hr NIRSpec/PRISM data. Color segments represent the range of the molecular bands selected in the spectra (H$_2$O-blue, CH$_4$-orange, CO-pink). \textit{Lower panel:} Max/min flux ratio of the two spectra shown in the upper panel. The horizontal dotted line represents the wavelengths where there is no variability.}
    \label{fig:spectrum_A}
\end{figure*}

We built a ``white light curve", which includes the flux of the full wavelength spectra (0.6 -- 5.3 $\mu$m), where we measure the brightest and faintest spectra corresponding to the peaks/troughs of this light curve. Figure \ref{fig:spectrum_A} (upper panels) shows the median of the 21 brightest spectra (or maximum flux) and the 21 faintest spectra (or minimum flux)for the A and B components (left and right panels, respectively). The brightest spectra of A object are found at $\sim$6.88~hr and at $\sim2.62$~hr for B object, and the faintest spectra are found at $\sim$ 1.47 hr for A object and $\sim4.17$~hr for B object, in the white light curve. Figure \ref{fig:spectrum_A} highlights the wavelength regions covered by each of the selected molecular bands: three water (H$_2$O-1, -2, -3) bands in the ranges of 1.36 -- 1.43 $\mu$m, 1.89~--~2.02~$\mu$m, and 2.68~--~3.00~$\mu$m (blue color), five methane (CH$_4$-1, -2, -3, -4, and -5) bands in the ranges of 1.10~--~1.19~$\mu$m, 1.65~--~1.75~$\mu$m, 2.19~--~2.28~$\mu$m, 3.10~--~3.28~$\mu$m, and 3.80~--~4.02~$\mu$m (orange color), and two carbon monoxide (CO-1, and -2) bands in the ranges of 2.32~--~2.40~$\mu$m, 4.48~--~5.00~$\mu$m (pink color). 


The lower panel of Figure \ref{fig:spectrum_A} shows the maximum/minimum ratio (lower panel). We confirmed the wavelength dependence of the variability, which is also presented in \citealt{biller2024jwst}. WISE 1049A shows a weaker wavelength dependence and a much lower variability amplitude at $<2$ $\mu$m than the longer wavelength CO feature (Figure \ref{fig:spectrum_A}, bottom left panel). WISE 1049B (Figure \ref{fig:spectrum_A}, bottom right panel) shows higher maximum deviation, mainly for wavelengths between 1.0~--~2.5~$\mu$m, and this seems to decrease with increasing wavelength. In Section \ref{Sec:physicalmec}, we discuss in detail the physical mechanisms that could explain this variability at different wavelengths.




\subsection{Light Curves within Molecular Absorption Features} \label{sec:lightcurves}

\begin{figure*}
    \centering
    \includegraphics[width=0.9\linewidth]{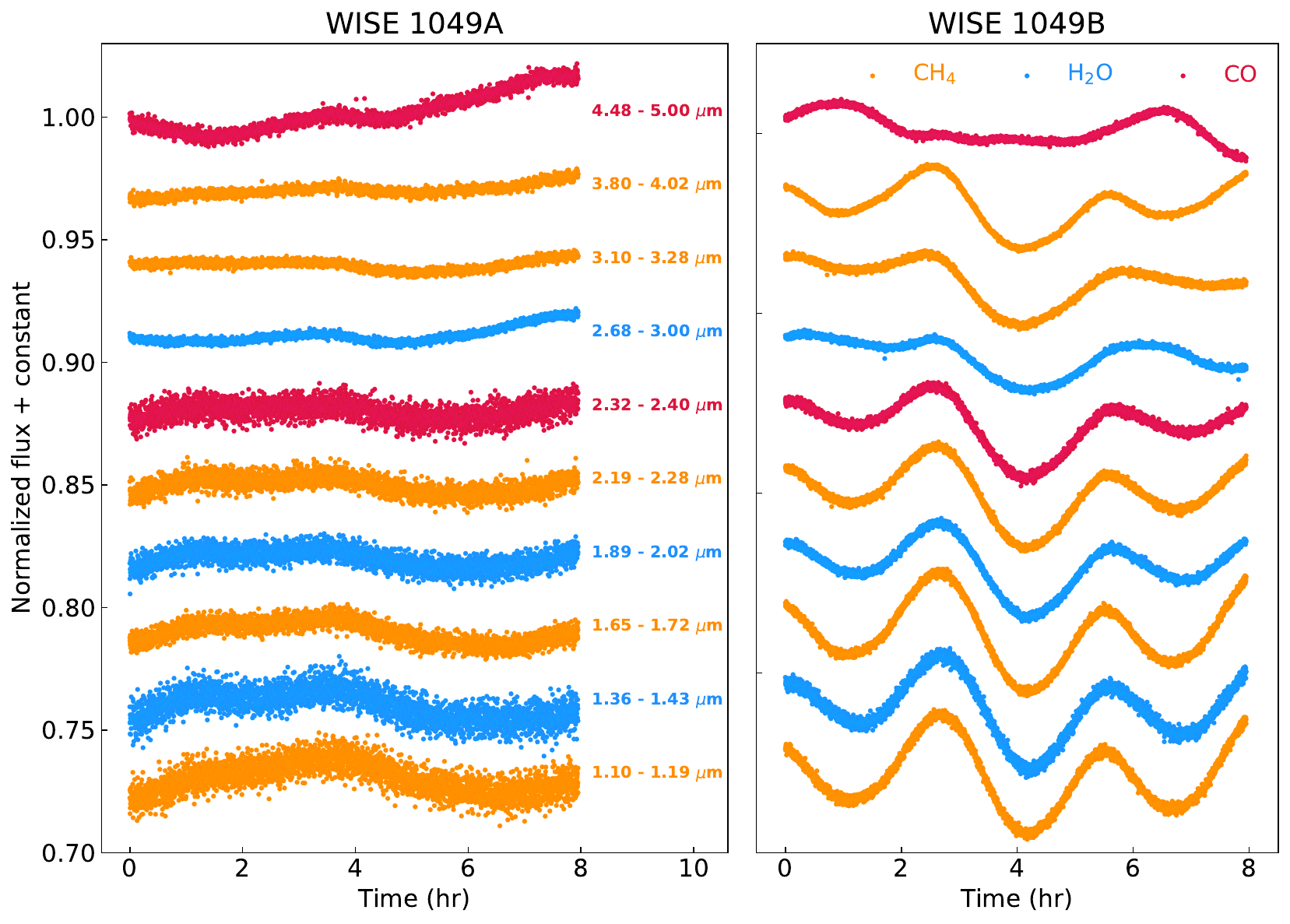}
    \caption{WISE 1049A (\textit{left}) and WISE 1049B (\textit{right}) light curves for each molecular band selected of H$_2$O - blue, CH$_4$ - orange, CO - pink, sorted by wavelength (top -- 0.6 $\mu$m and bottom 5.3 $\mu$m). }
    \label{lc_AB}
\end{figure*}

We selected the wavelength ranges covering the molecular bands of H$_2$O, CH$_4$, and CO, based mainly on the results of previous studies of objects with similar spectral types (\citealt{cushing2005infrared}, \citealt{marley2021sonora}, \citealt{Miles_2023}, \citealt{biller2024jwst}). In cases where molecules overlapped, we cut these ranges to retain regions where only one molecule contributed. In addition, we confirmed our wavelength ranges to trace unambiguous pressure levels of the atmosphere, according to the Contribution Function discussed in Section \ref{Sec:contribution_function} and Appendix \ref{AppendixC}. We measured the time-dependent flux for WISE 1049AB inside the following molecular bands:  H$_2$O-1 (1.36~--~1.43~$\mu$m),  H$_2$O-2 (1.89~--~2.02~$\mu$m), and H$_2$O-3 (2.68~--~3.00~$\mu$m), CH$_4$-1 (1.10 -- 1.19 $\mu$m), CH$_4$-2 (1.65~--~1.75~$\mu$m), CH$_4$-3 (2.19~--~2.28~$\mu$m), CH$_4$-4 ( 3.10~--~3.28~$\mu$m), and CH$_4$-5 (3.80~--~4.02~$\mu$m), CO-1 (2.32~--~2.40~$\mu$m), and CO-2 (4.48~--~5.00~$\mu$m). The molecular bands of CH$_4$ and H$_2$O in the ranges of 1.10 -- 1.19 $\mu$m and 1.36~--~1.43~$\mu$m, respectively, make it impossible to separate the molecular features. However, this does not represent a problem for our analysis, as they explore deep areas of the atmosphere where heterogeneous clouds dominate variability. We averaged every 16 spectra (time-sampling $\sim$7.5 s) to improve the signal-to-noise of the light curve and integrated the flux over the wavelength ranges. Then we normalized them by their median values. We present the light curves for WISE 1049A and WISE 1049B in Figure \ref{lc_AB}. Each data point is the integral of each spectrum in the specific wavelength range of the molecular feature only. We remove data outliers, identifying them as three standard deviations outside the trend of the surrounding data. 

In Figure \ref{lc_AB}, the light curves show the variability within each H$_2$O, CH$_4$, and CO molecular bands for both WISE 1049A and B, which have different shapes and maximum deviations. In the following sections, we will analyze these light curves, including their amplitudes, periods, and phases, and discuss the possible causes of the variability observed for each molecular band.

\subsubsection{Light curves correlation by molecular features}\label{sec:lc-correlations}
Figure \ref{lc_AB} reveals a visual correlation between the light curves inside the wavelength ranges where the molecules contemplated in this work are. We quantify the possible correlation between the light curves of all the molecular features of H$_2$O, CH$_4$, and CO. For this, we calculate Kendall's correlation coefficient since this coefficient can discern strictly monotonic correlations, whether linear or non-linear, and is robust to outliers and small sample sizes. We use the package \texttt{scipy.stast.kendalltau} to calculate for all combinations of light curves of the three molecules. For both objects WISE 1049 A and B, we found a higher correlation between CH$_4$ vs H$_2$O ($\tau\sim 0.7$, for WISE 1049B) than CH$_4$ vs CO, or H$_2$O vs CO ($\tau\sim 0.4$, for WISE 1049B), which is the expected behavior according to equation \ref{eq:reaction}, provided that the chemical equilibrium assumption is satisfied. However, in WISE 1049A, we found weak correlations in all three cases, which could indicate that this chemical balance is not necessarily being maintained.. We found that the correlation between CH$_4$ vs H$_2$O is $\tau\sim 0.3$ and the correlations between CH$_4$ vs CO or H$_2$O vs CO are $\tau\sim 0.1$.

\subsection{Light Curves Fitting}\label{sec:fittingLC}

To identify the main characteristics of each light curve, we fitted each normalized light curve with a sinusoidal model, which represents a planetary-scale wave, following the analysis process to \citet{Apai2021} and \citet{fuda2024latitude}. We varied the function between one and six ($n$) sine waves in the form: 
\begin{equation}
    1 + \sum_{i=1}^{n} A_i \sin\left(\frac{2\pi}{P_i} + \phi_i\right).
\end{equation}

We obtained a total of between 3 and 18 parameters for fitting. We used the \texttt{curve$\_$fit} package in \texttt{scipy} \citep{2020SciPy-NMeth} to find the best fits to our light curves. For WISE 1049B, we used the parameters obtained by \cite{fuda2024latitude} in their periodic analysis of the system as initial priors to start our fitting process of WISE 1049B, since its analysis involves combined light curves of both objects with a higher contribution from WISE 1049B. However, for WISE 1049A, we changed the initial guess of the period ($\sim$7 hr \citealt{Apai2021}). In addition, we limited the range of parameters as follows: 0.01 -- 10\% for the amplitude of both objects, 1~--~10~hr for the period of WISE 1049B, and 1~--~15~hr for the period of WISE 1049A. We varied the number of sinusoidal functions, and we made use of three different statistical methods to find the number of functions that best fit the light curve data for each molecular band.

We used two statistical criteria to find the best fit, based on non-nested models, the Akaike Information Criterion (\textit{AIC}), and the Bayesian Information Criterion (\textit{BIC}). These methods help us balance model complexity, i.e, the number of parameters, with goodness of fit, which is how well the model explains the data. However, each method provides us with different weights for these two important fitting characteristics. In the following subsections, we describe these two methods, all of which find the same number of sinusoidal functions as the best fit.

\begin{figure*}
    \centering
    \includegraphics[width=1\linewidth]{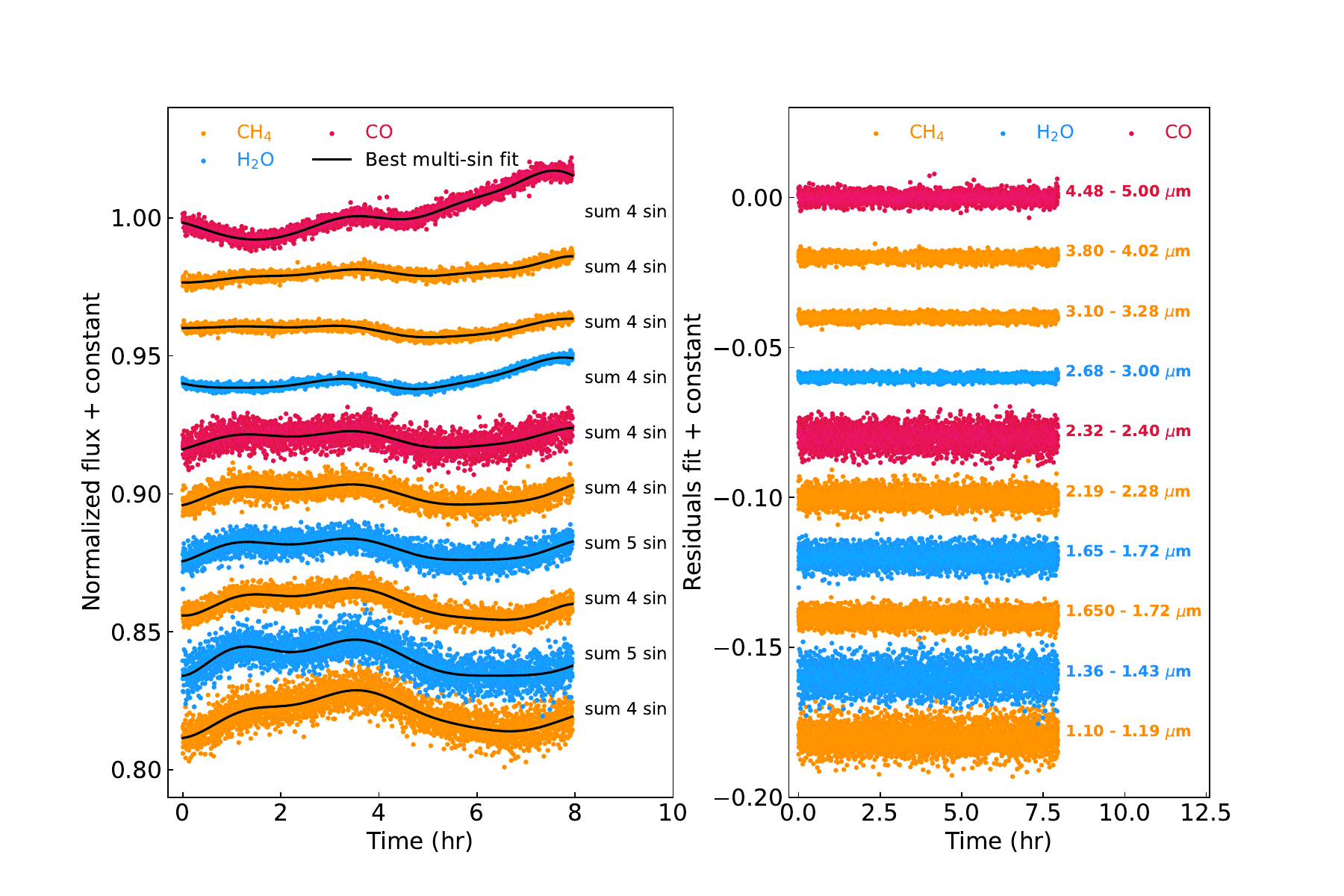}
    \caption{\textit{Left panel:} WISE 1049A light curves for different molecular bands of H$_2$O - blue, CH$_4$ - orange, CO - pink, each with the best fit (black), next to each is the number of sinusoidal functions found as seen in the best fit. \textit{Right panel:} the residual fit (data minus fit).}
    \label{fig:lc-A}
\end{figure*}

\begin{figure*}
    \centering
    \includegraphics[width=1\linewidth]{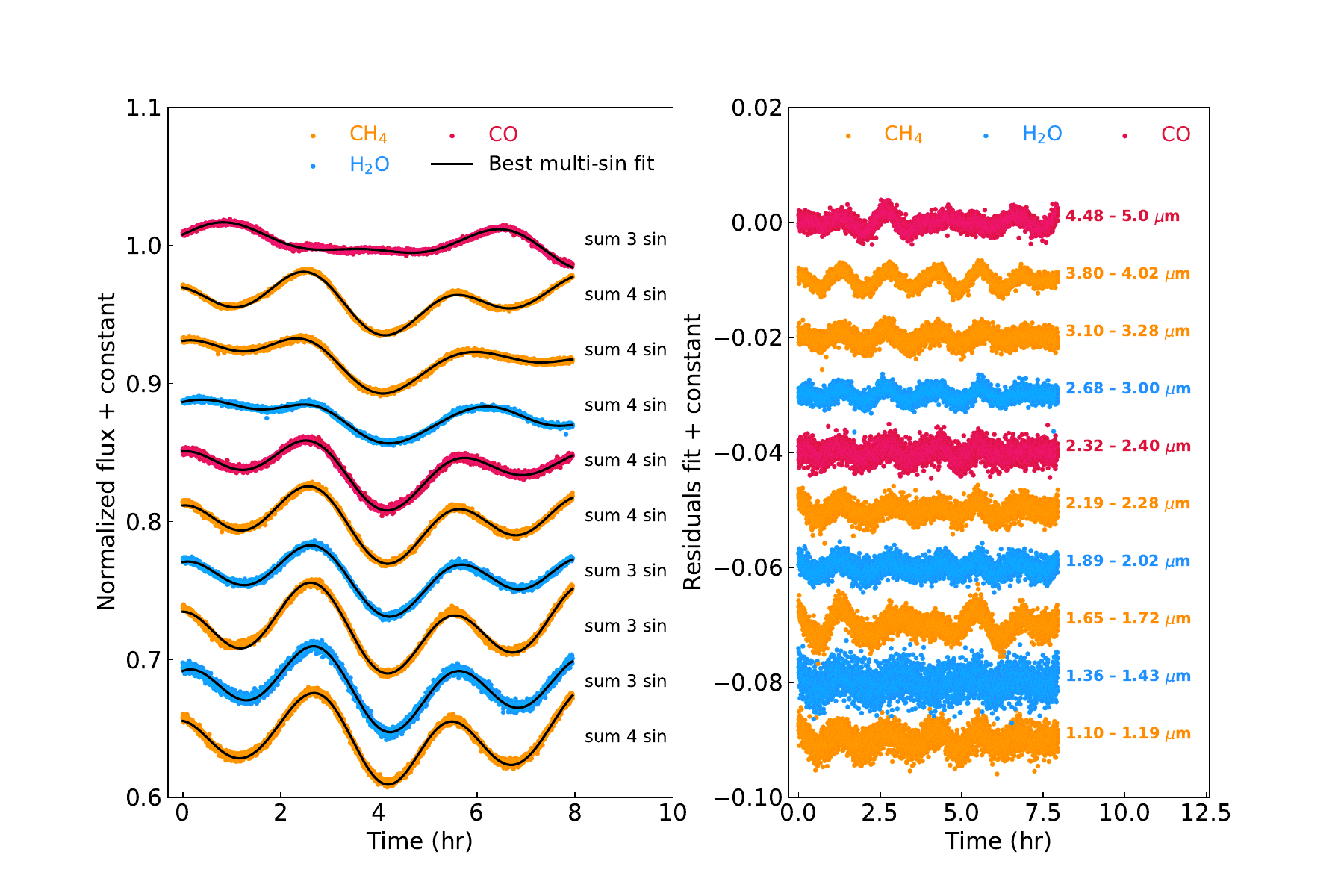}
    \caption{\textit{Left panel:} WISE 1049B light curves for different molecular bands of H$_2$O - blue, CH$_4$ - orange, CO - pink, each with the best fit (black), next to each is the number of sinusoidal functions found as seen in the best fit. \textit{Right panel:} the residual fit (data minus fit).}
    \label{fig:lc-B}
\end{figure*}

\subsubsection{Akaike Information Criterion (AIC)}

To quantitatively compare the quality of different fits to our light curves, we used the Akaike Information Criterion (AIC), which estimates the relative amount of information lost when a given model is used to represent the data. Lower AIC values indicate models that better balance goodness of fit and model complexity.

Following the formal definition for Gaussian likelihoods (eq 13 of \citealt{thorngren2025bayesian}), the log-likelihood is expressed as:
\begin{equation}
\ln (L)= -\frac{1}{2}\chi^2 - \frac{1}{2} \sum_i 
\ln (2\pi \sigma_i^2),
\label{eq:likelihood}
\end{equation}

where 
$\chi^2 = \sum_i \frac{(y_i - f_i)^2}{\sigma_i^2}$, $y_i$ are the observed fluxes, $f_i$ the model predictions, and $\sigma_i$ the corresponding uncertainties.

The AIC \citep{akaike2011akaike} is defined as $$AIC=2 k -2\ln(L^2),$$ where $k$ is the number of free parameters in the model, \textbf{and $L$ is} the maximum likelihood of the model, defined in the equation \ref{eq:likelihood}, implemented from the \texttt{scikit-learn} package. This expression properly accounts for the data uncertainties 
$\sigma_i$. This implies the first term penalizes the model that has more parameters $k$ to avoid overfitting, and the second term measures the goodness of fit. The lower AIC values indicate the best model.  ensuring that the likelihood normalization is consistent across models. In practice, differences in AIC values ( $\Delta AIC$) between models indicate the relative strength of evidence, with $\Delta AIC>10$ typically considered strong evidence against the higher-AIC model.

\subsubsection{Bayesian Information Criterion (BIC)}
We also evaluated the model complexity using the Bayesian Information Criterion \citep{neath2012bayesian} (BIC), which introduces a stronger penalty for the number of parameters relative to sample size. The BIC is defined as:
\begin{equation*}
    BIC = k\ln(N) - 2 \ln (L),
\end{equation*}
where $N$ is the number of data points and $L$ the likelihood defined above.

The first term adds a stronger penalty for complexity compared to \textit{AIC}, as the number of observations increases. This means \textit{BIC} is more conservative in allowing model complexity than \textit{AIC}. Like \textit{AIC}, the second term evaluates how well the model fits the data. For this method, the lower values of BIC indicate the best model. Compared to AIC, BIC tends to favor simpler models, especially with larger datasets. The BIC thus provides a conservative assessment of whether adding more parameters (e.g., additional sinusoidal components) is justified by a significant improvement in fit quality. Models with lower BIC are preferred, with $\Delta BIC>10$ considered strong evidence against the higher-BIC model.

\subsection{Identifying the best fits for WISE 1049A and B light curves}\label{structures_map}

We fitted between one and six sinusoidal functions to each light curve. We found that for most molecular bands in both objects, the three statistical methods identified the same number of sinusoidal tasks as the best fit. If two methods do not provide the same number of sinusoidal functions as the best fit, we conduct a visual inspection. If the difference between the values of the methods is not significant, we select the fit with the fewest sinusoidal functions to avoid overfitting. Table \ref{Table:ParametersFitA} in Appendix \ref{fits} summarizes all the parameters found for the best fit of object A. The parameters of the best fit of object B are summarized in the Appendix \ref{fits}, Table \ref{Table:ParametersFitB}.

We found that the best fit for all the light curves of the molecular bands present in WISE 1049A is four sinusoidal functions (see Figure \ref{fig:lc-A}, left).  Furthermore, the right panel of Figure \ref{fig:lc-A} shows the residuals between the fit and the observational data. These residuals are rather flat, and they do not follow any pattern other than a gaussian dispersion around zero.
From the amplitudes obtained of the best fits (Appendix \ref{fits} - Table \ref{Table:ParametersFitA}), we detect that at least one wave have a higher value than the others, which allows us to infer that most of the variability is produced by a dominant band in the variability we observe. In addition, at least one of the sinusoidal functions has a period between 6--8 hr, which is close to the period calculated by \citet{Apai2021} with a $P_{rot} = 7$ hr ($k=1$ waves, where $k$ is the wavenumber waves). However, some harmonics can also be seen with a period between 3--3.5 hr ($k=2$ waves). It is important to clarify that in the case where we find periods longer than expected by the literature, such as the case of H$2$O where we found a period of 12.5 hr, it is necessary to be careful because this result is beyond the baseline time of our data, so we cannot assign a physical meaning to it.

For WISE 1049B, we find the best fit between three and four sinusoidal functions (Left panel - Figure \ref{fig:lc-B}). 
The right panel in Figure \ref{fig:lc-B} shows the residuals between the fits and the light curves of WISE 1049B; we observe that the residuals are not flat, and furthermore, they appear to present a periodicity. The periodicity remains unresolved even when the number of sinusoidal functions is increased or decreased, suggesting that WISE 1049B has a significantly more complex atmospheric structure than WISE 1049A, which could contribute to its higher amplitude variability.
We found that the amplitude of at least one function of each molecular band has an amplitude value between 2 -- 5\%, close to the maximum deviation reported, while the other complementary waves have much smaller amplitudes (summarized in the Appendix - Table \ref{Table:ParametersFitB}). In addition, at least one of the sinusoidal functions has a period between 4.8 -- 5.8 hr, which is close to periods reported by \citet{Apai2021} and \citet{fuda2024latitude} with a $P_{rot} = 5$ hr mark ($k=1$ waves), and some harmonics with period around 2.5 hr ($k=2$ waves).

\subsection{Periods}

\begin{figure}
    \centering
    \includegraphics[width=1\linewidth]{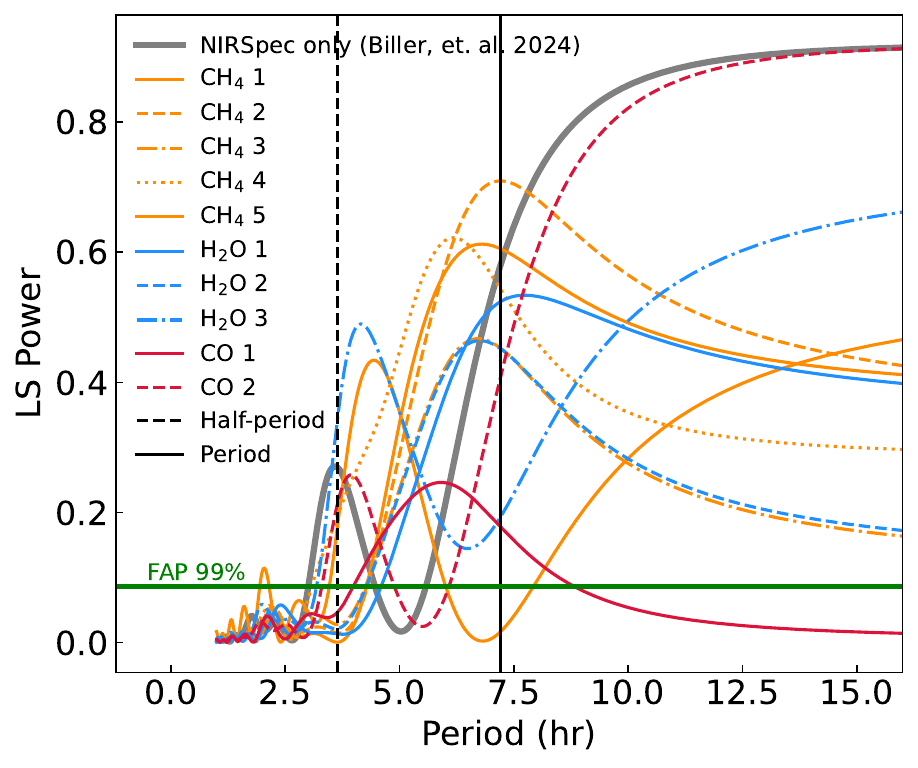}
    \caption{Lomb-Scargle Periodogram of WISE 1049A. The vertical lines represent the half-period at $3.46$ hr (dashed) and the period at $6.94$ hr (solid) reported by \cite{Apai2021}. Each color and line style represents a different range of molecular bands selected. In addition, we add the periodogram (grey) calculated with the `only NIRSpec light curve', also presented in \citet{biller2024jwst}.}
    \label{fig:periodogram-A}
\end{figure}

\begin{figure}
    \centering
    \includegraphics[width=1\linewidth]{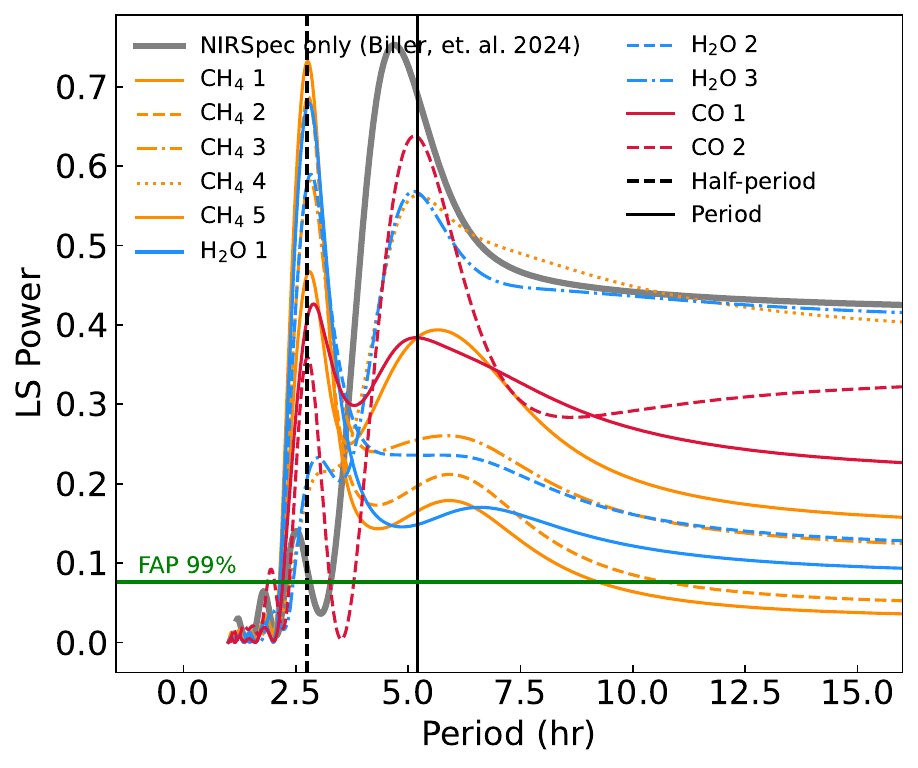}
    \caption{Lomb-Scargle Periodogram of WISE 1049B. The vertical lines represent the half-period at $2.5$ hr (dashed) and the period at $5.28$ hr (solid) reported by \citet{Apai2021}. Each color and line style represents a different range of molecular bands selected. In addition, we add the periodogram (grey) calculated with the `only NIRSpec light curve', also presented in \citet{biller2024jwst}.}
    \label{fig:periodogram-B}
\end{figure}

To confirm the rotational periods obtained in Section \ref{structures_map}, we run a Lomb-Scargle periodogram method \citep{vanderplas2018understanding} using the \texttt{astropy.timeseries.LombScargle} python package, which is a common method in the frequency analysis of unequally spaced data, equivalent to least-squares fitting of sine waves. We derived the periodograms from each of the light curves for each molecular band of WISE 1049A and B (as shown in Figure \ref{lc_AB}). Figures \ref{fig:periodogram-A} and \ref{fig:periodogram-B} show the periodograms for all of the molecular bands selected from WISE 1049A and B, respectively. We considered periods ranging from 1.2 to 16 hours, as in \citealt{biller2024jwst} and \citealt{chen2025jwst}, and calculated the 99 percent False Alarm Probability (FAP) power using the built-in bootstrap method.

For WISE 1049A, the light curves cover about one rotational period of the object ($6.94$ hr, \citealt{Apai2021}). We found that most light curves have significant power at periods $\geq$6 hours, suggesting that the true period of the A component is longer than 6 hours, which coincides with the period reported by \cite{Apai2021}. In addition, we found a peak with a smaller power around 3 to 5 hr, which could be related to the half-period. Thus, based on our period analysis, we cannot confirm that the period of WISE 1049A is $\sim7$~hr. We obtain similar results for all the molecular light curves as the periodograms presented by \citealt{biller2024jwst} for the ``NIRSpec light curves", where significant power remains at periods greater than 10 hours. In addition, we can infer that it is longer than the period of WISE 1049B, and longer monitoring is needed to confirm this.

In the case of WISE 1049B, from the periodograms, we measured a period of $5.14$ hr, using the maximum of the CO molecular band in the 4.3--5.0 $\mu$m range, consistent with various studies (e.g. \citealt{burgasser2014monitoring}, \citealt{buenzli2015cloud}, \citealt{Apai2021}, \citealt{fuda2024latitude}, \citealt{biller2024jwst}). We found this same peak with lower power at around 5 hr for the other molecular bands. This confirms the periods calculated with the best fitting, in Section \ref{structures_map}, where the sinusoidal functions with higher amplitude have periods around $5$~hr and 2.5~hr. In addition, we notice a second significant peak in all molecular bands at $2.76$~hr (also found in the periodogram of the NIRSpec light curve of \citealt{biller2024jwst}). This second peak could be directly related to a half-period of the principal peak. Although for the majority of H$_2$O and CH$_4$ molecular bands, this appears to be the stronger peak, we associate this peak with a harmonic of the main period, but the physical reason why this happens is unclear.

The periods calculated with the periodograms presented in this Section agree with the periods obtained from the fits to our light curves. At least one wave of the fits finds periods close to the rotation periods of the literature ($\sim$7~hr for WISE 1049A and $\sim$5 hr for WISE 1049B), and some of the other waves are a harmonic of the rotation period.

\subsection{Maximum deviation of the Variability}\label{amplitudes}
We used the definition of the maximum deviation presented in \citet{biller2024jwst} as the percent of relative variability between the highest peak and deepest trough observed in each molecular feature's light curves. For each of the light curves, we measured the maximum deviation from the average. Specifically, we subtracted the mean flux of 21 spectra around the maximum and minimum peaks in each light curve, multiplied the result by 100, and then divided it by the average of the light curve data. We also compared the maximum deviation calculated using the fits of each light curve by subtracting the maximum and minimum values in the peak of each light curve, multiplying by 100, and then dividing by the average of the light curve fit data. Both methods yielded similar results. All amplitudes for each molecular band and each object, with both methods, are summarized in Table \ref{table:limits_molecules}.

WISE 1049A shows peak-to-peak maximum deviation between 0.70 -- 2.66~\%, where CH$_4$-4 (3.10 -- 3.40~$\mu$m) has the smallest amplitude and CO-2 (4.48 -- 5.00 $\mu$m) has the highest amplitude. WISE 1049B shows maximum deviations between 3.43 -- 7.00~\% where CO-2 (4.30 -- 5.00 $\mu$m) shows the smallest amplitude and CH$_4$-1 (1.10 -- 1.19 $\mu$m) has the highest amplitude. The uncertainties of the amplitudes are calculated following the same procedure presented in \cite{biller2024jwst}, where initially, the standard deviation ($\sigma_{pt}$) defined in \citealt{radigan2014strong} for each value of the light curves is measured. Thus, we propagate the error and calculate the standard deviation of the flux at the 21 maximum and minimum values of the peaks in the observational data of the light curves. These uncertainties are around $\sim 0.10\%$, consistent with the results of \cite{biller2024jwst} and \cite{chen2025jwst}. They are included in the Table \ref{table:limits_molecules}, but we do not quote the uncertainties in the rest of the text. In addition, we calculated the uncertainties for the maximum deviation using the fitted light curves, according to the numerical Jacobian obtained for the covariance matrix of the parameters of the best-fit. The fit uncertainties are smaller in comparison to the observational data values because the model is being conditioned on the full light-curve instead of the spread of the 21 minimum/maximum values of the flux. We note that the maximum deviations derived from the models are systematically lower than those measured from the observations. This difference is expected because the observed light curves contain photometric noise that broadens the measured flux extrema, while the fitted models are noise-free and represent the mean signal. However, the good agreement between observed and modeled light curves (Figures \ref{fig:lc-A} and \ref{fig:lc-B}) indicates that this effect is not a systematic bias of our method but a natural consequence of noise smoothing in the models. It is worth noting that we also compare the maximum deviation of the selected molecular bands and their pseudocontinuum to corroborate our analysis. This discussion is presented in Appendix~\ref{AppendixB}.

\section{Depth-dependent Variability}\label{sec:depth_dependent_variability}

\subsection{Tracing different Atmospheric levels of WISE 1049A and B}\label{Sec:contribution_function}

We used the radiative transfer code \texttt{picaso} \citep{Mukherjee_2023, Batalha_2019} to obtain the depths/pressure levels of the atmosphere of WISE 1049AB traced by each of the molecular bands for which we measured the variability. To obtain those depths/pressure levels traced, we used the contribution function provided by the best-fitting model to WISE~1049AB, which describes how much each layer of a brown dwarf’s atmosphere contributes to the emergent radiation at a specific wavelength. We calculated the contribution function for atmospheres with $T_\mathrm{eff}$= 1300, 1400, and 1500 K and $\log g$ = 5.0 [cgs] for clear atmospheres and cloudy atmospheres with $f_\mathrm{sed}$= 2 (we discussed the main differences in the Appendix \ref{AppendixC}). For clear atmospheres, we used the temperature-pressure profiles from Sonora Bobcat \citep{Marley_2021}. For cloudy atmospheres, we used those from Sonora Diamondback \citep{Morley_2024}, including homogeneous clouds, as MgSiO$_3$, Mg$_2$SiO$_4$, Al$_2$O$_3$, MnS, Na$_2$S. We used a custom-made opacity database created using data from \citet{lupu_2022} at resolution, $R$, of 500,000. We ran our models at R = 5,000 and downsampled to the resolution of our NIRSpec/PRISM spectra ($R \sim 300$). The \texttt{picaso} code calculates the thermal contribution function following \citet{Lothringer_2018}. 

\begin{figure}[h]
    \centering
    \includegraphics[width=1.02\linewidth]{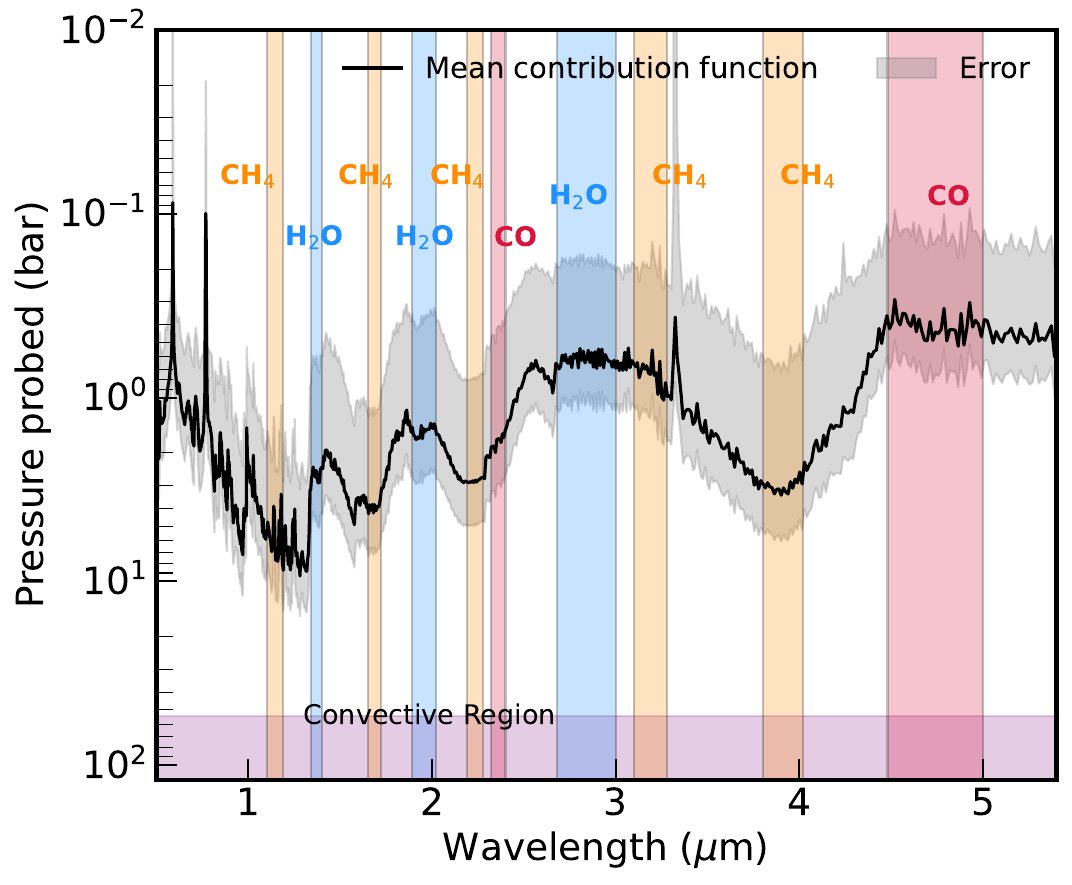}
    \caption{Contribution function from 0.6-5.3 $\mu$m Sonora Diamondback model with $T_{\text{eff}} = 1300$K, $f_{\text{sed}}=2$, and equilibrium chemistry. The black line indicates the pressure at which the maximum flux is emitted. The grey zone represents the pressures from where 80\% to 95\% of the flux is emitted, based on the weighted pressure information from our contribution function \citep{Lothringer_2018}.}
    \label{fig:cont_func}
\end{figure}

We traced the average pressure probed by each molecular band by identifying the wavelength at which the contribution function peaks. We used a cloudy contribution function for a $T_\mathrm{eff}$=~1300~K, log~$g$~=~5 brown dwarf, assuming equilibrium chemistry, as derived by \citealt{biller2024jwst} using evolutionary models. We discuss in detail the selection of the best $\mathrm{T_{eff}}$ and log~g to derive the contribution function in the Appendix \ref{AppendixC}. The chosen contribution function is shown in Figure \ref{fig:cont_func}. 
As shown in Figure \ref{fig:cont_func}, our observations probe a wide range of pressures, from $\sim$10~bar to a few hundred mbar, tracing most of the atmosphere of WISE 1049AB. We note that the contribution functions are representative of a horizontally homogeneous column with the same properties ($T_\mathrm{eff}$, $\log g$, etc.) as the best fit to the average brown dwarf spectrum. In reality, the atmospheres of brown dwarfs like WISE 1049AB are known to be heterogeneous, so our pressures are indicative of the average pressure probed by our observations.

\begin{deluxetable*}{cccccccc}
\tablecaption{Wavelength ranges of molecular bands studied here with the pressures that they trace, and the maximum deviation found for each of them. References of the ranges: (1) \citet{cushing2005infrared}, (2) \cite{marley2021sonora}, (3) \cite{Miles_2023}, (4) \cite{biller2024jwst}.} \label{table:limits_molecules}
\tablehead{
\colhead{Molecular} & \colhead{Wavelength range} & \colhead{Pressure traced} &
\multicolumn{2}{c}{Max deviation$^b$} &
\multicolumn{2}{c}{Max deviation$^b$} &
\colhead{Range$^c$} \\
\colhead{band} & \colhead{($\mu$m)} & \colhead{(Bar)$^a$} &
\multicolumn{2}{c}{WISE 1049A (\%)} &
\multicolumn{2}{c}{WISE 1049B (\%)} &
\colhead{references} \\
\colhead{} & \colhead{} & \colhead{} &
\colhead{Obs data} & \colhead{fit} &
\colhead{Obs data} & \colhead{fit} &
\colhead{}
}
\startdata
CH$_4$ 1           & 1.10 -- 1.19   & $6.01\pm 1.01$ & 2.06$\pm$ 0.12 & 1.96$\pm$0.02 & 7.00$\pm$ 0.08 & 6.62$\pm$ 0.03  & (1) \\
H$_2$O 1           & 1.36 -- 1.43   & $2.66\pm 0.93$ & 1.58$\pm$ 0.12 & 1.36$\pm$0.02 & 6.30$\pm$ 0.09 & 6.23$\pm$ 0.01  & (1), (3), (4)   \\
CH$_4$ 2           & 1.65 -- 1.72   & $3.95\pm 0.65$ & 1.35$\pm$ 0.07 & 1.12$\pm$0.02 & 6.79$\pm$ 0.06 & 6.57$\pm$ 0.01  & (1) \\
H$_2$O 2           & 1.89 -- 2.02   & $1.53\pm 0.31$ & 1.25$\pm$ 0.09 & 1.14$\pm$0.01 & 5.75$\pm$ 0.05 & 5.19$\pm$ 0.01  & (1), (3), (4)   \\
CH$_4$ 3           & 2.19 -- 2.28   & $2.83\pm 0.52$ & 1.30$\pm$ 0.11 & 0.98$\pm$0.01 & 5.72$\pm$ 0.06 & 5.64$\pm$ 0.02  & (4)     \\
CO 1               & 2.32 -- 2.40   & $1.73\pm 0.27$ & 0.81$\pm$ 0.10 & 0.75$\pm$0.02 & 5.21$\pm$ 0.07 & 5.09$\pm$ 0.02  & (3), (4)  \\
H$_2$O 3           & 2.68 -- 3.00   & $0.63\pm 0.12$ & 1.09$\pm$ 0.06 & 0.81$\pm$0.01 & 3.49$\pm$ 0.08 & 3.18$\pm$ 0.02  & (1), (3), (4) \\
CH$_4$ 4           & 3.10 -- 3.28   & $0.76\pm 0.18$ & 0.70$\pm$ 0.06 & 0.69$\pm$0.01 & 4.40$\pm$ 0.06 & 4.00$\pm$ 0.01  &(1), (3), (4) \\
CH$_4$ 5           & 3.80 -- 4.02   & $3.06\pm 0.51$ & 1.24$\pm$ 0.06 & 0.98$\pm$0.01 & 4.85$\pm$ 0.05 & 4.61$\pm$ 0.02  &(2), (3) \\
CO 2               & 4.48 -- 5.00   & $0.41\pm 0.09$ & 2.66$\pm$ 0.07 & 2.44$\pm$0.02 & 3.43$\pm$ 0.07 & 3.25$\pm$ 0.01  & (3), (4) \\
\enddata
\tablecomments{$^a$Pressures traced for each molecular band as explained in Section \ref{Sec:contribution_function}.  $^b$The uncertainties of the relative amplitudes for the observational data (\%) are associated with error propagation of the standard deviation of individual value in the light curves and the standard deviation of the flux around the maximum, minimum, and for the fit values are associated with the uncertainty band of a fitted model using numerical Jacobian.$^c$ The wavelength ranges were chosen to avoid regions with significant changes in opacity.}
\end{deluxetable*}

\subsection{Physical mechanisms of variability}\label{Sec:physicalmec}

In this Section, we discuss how different physical mechanisms might explain the variability observed at the wavelengths shown in Table~\ref{table:limits_molecules} that trace distinct pressure levels of the atmosphere of WISE~1049AB.

\begin{figure*}
    \centering
    \includegraphics[width=1\linewidth]{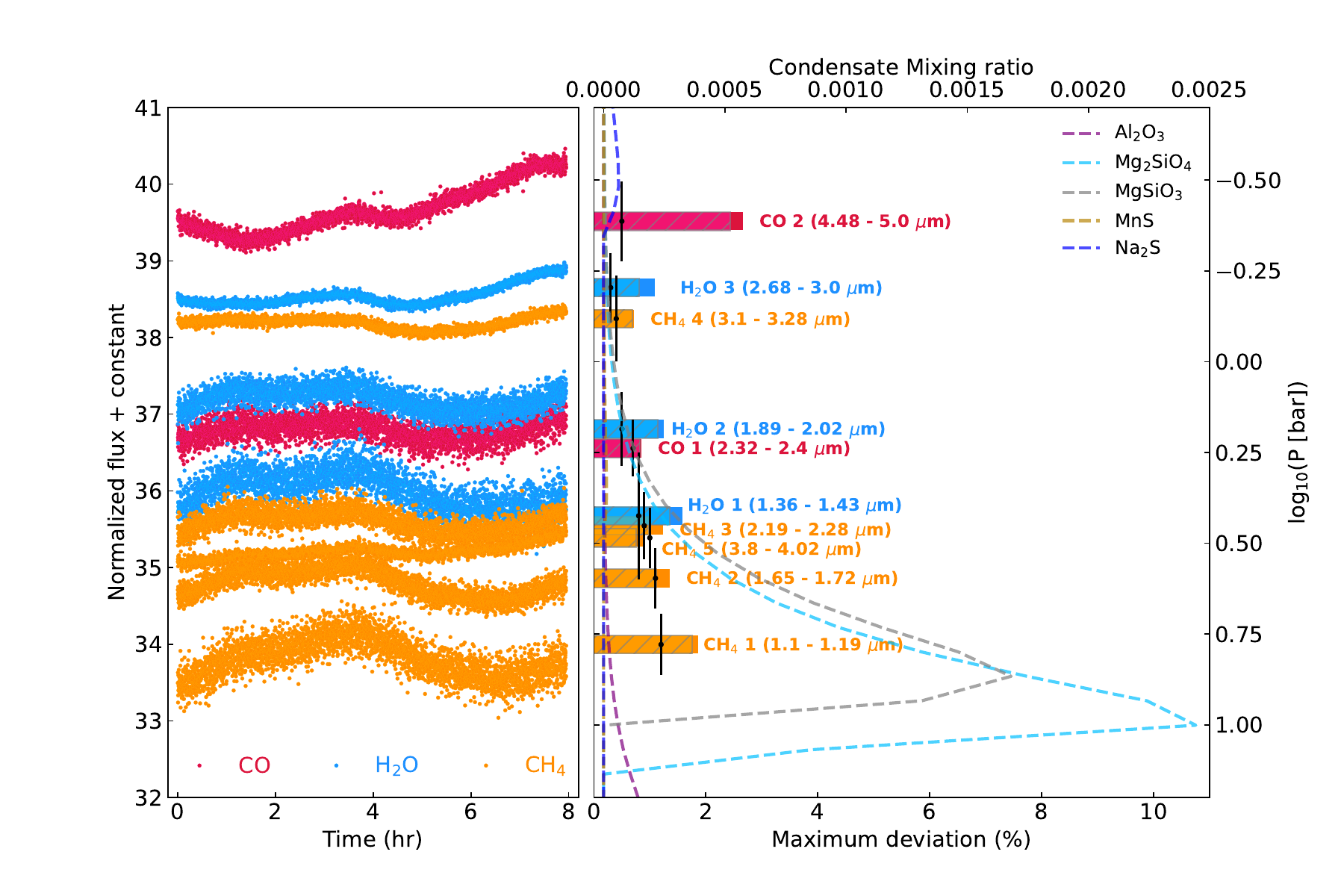}
    \caption{\textit{Left panel:} Light curves of selected molecular bands in WISE 1049A spectra, organized by pressure level. \textit{Right panel:} Mole fraction of the silicate clouds expected for a $\mathrm{T_{eff}}$=1300 K brown dwarf (dashed colored lines). The horizontal bars represent the maximum deviation, obtained from the raw light curve data (filled bar) and from the fit data (bar transparent with diagonal lines), in the position of the light curves obtained by the contribution function. The uncertainties of the maximum deviations are listed in the Table \ref{table:limits_molecules}. Next to it, we display the name of the molecular band and the wavelength range it is tracing.}
    \label{fig:mechanism-A}
\end{figure*}

\begin{figure*}
    \centering
    \includegraphics[width=1\linewidth]{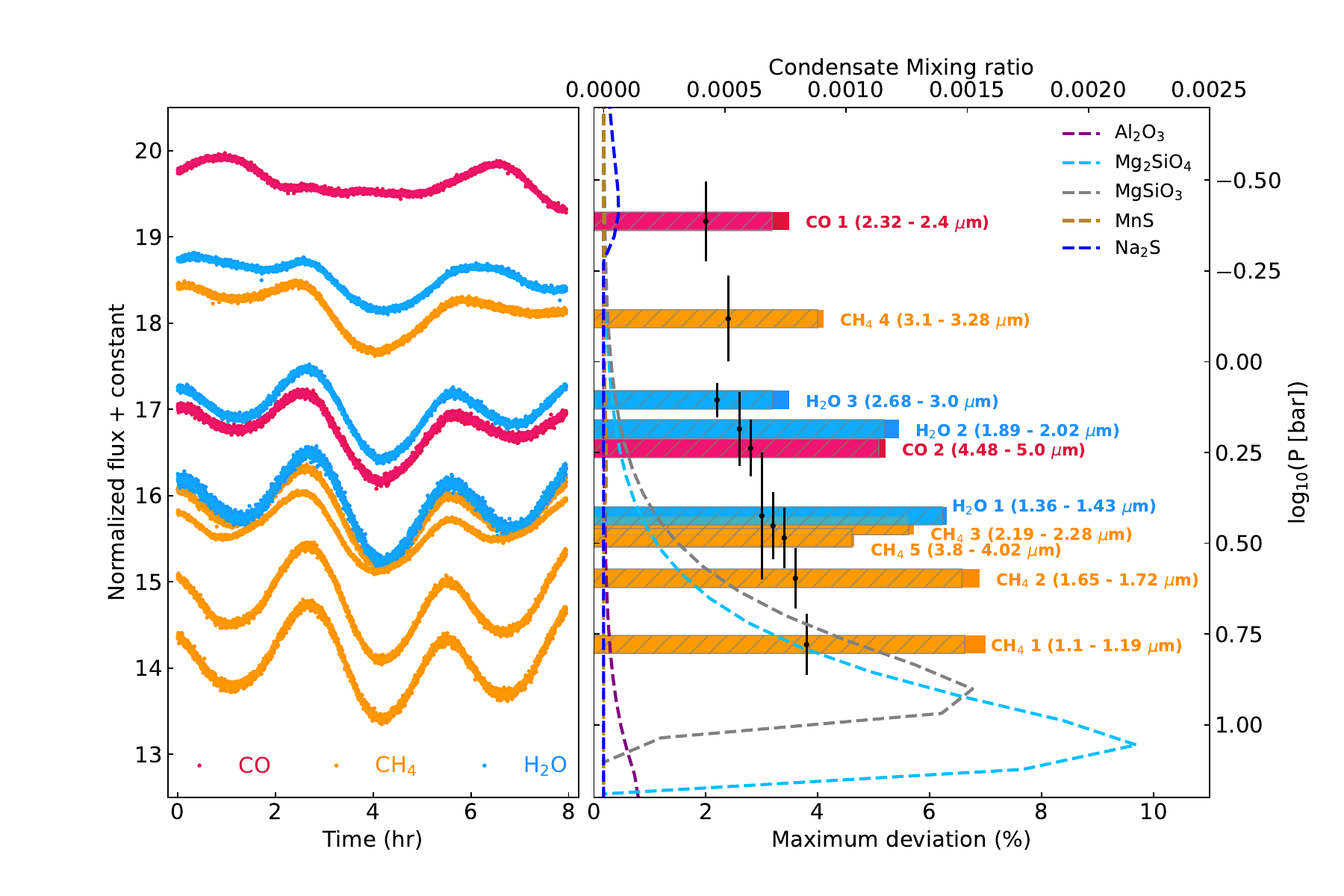}
    \caption{\textit{Left panel:} Light curves of selected molecular bands in WISE 1049B spectra, organized by pressure level. \textit{Right panel:} Mole fraction of the clouds expected for a $\mathrm{T_{eff}}$=1200 K brown dwarf (dashed colored lines). The horizontal bars represent the position of the light curves obtained by the contribution function and the maximum deviation, obtained from the raw light curve data (filled bar) and from the fit data (bar transparent with diagonal lines). The uncertainties of the maximum deviations are listed in the Table \ref{table:limits_molecules}. Next to it, we display the name of the molecular band and the wavelength range it is tracing.}
    \label{fig:mechanism-B}
\end{figure*}

In the left panel of Figures \ref{fig:mechanism-A} and \ref{fig:mechanism-B}, we show the light curves of WISE~1049~A and B, respectively, corresponding to each of the molecular bands specified in Table \ref{table:limits_molecules} ordered by the pressure levels as given by the contribution function. In the right panels of Figures \ref{fig:mechanism-A} and \ref{fig:mechanism-B}, we also show the condensate mixing ratio (mole fraction) of the different cloud species expected according to \cite{Marley2013}, in the atmospheres of WISE~1049~AB (dashed colored lines). The maximum deviations measured for each molecular band are plotted as colored horizontal bars. The solid color bars correspond to the maximum deviations calculated from the individual light curves. The overplotted dashed bars correspond to the maximum deviation as measured from the sinusoidal fits (see Section \ref{sec:fittingLC}, and Tables \ref{Table:ParametersFitA} and \ref{Table:ParametersFitB} of the Appendix for best fit parameters.)


Both WISE~1049 A and B show high variability in the light curves that trace their deep atmosphere $\sim$6 bar (the  CH$_4$-1  molecular feature). The CH$_4$-1  molecular feature light curve (1.10 -- 1.19 $\mu$m), which traces the 6.01~bar level, shows a 2.06 \% of maximum deviation for WISE 1049A and a 7.00 \% for WISE 1049B. 
The atmospheric models predict the highest condensate mixing ratio for the $\text{Mg}_2\text{SiO}_4$ and $\text{Mg}\text{SiO}_3$ silicate clouds at the pressure levels traced by the CH$_4$-1 molecule, which might explain why the highest variability is measured in this wavelength range.

However, it is worth highlighting two important points: 

\begin{enumerate}
    \item Different molecular features that trace similar pressure levels in the atmosphere could report different maximum deviations. For both objects, this can be observed in Figure \ref{fig:mechanism-A} and \ref{fig:mechanism-B}, right panels, for the $\mathrm{CH_{4}}$-3, $\mathrm{CH_{4}}$-5 and $\mathrm{H_{2}O}$-1 molecular features, which trace a pressure level of $\sim$3~bar. Similarly for the CO-1 and the $\mathrm{H_{2}O}$-2 molecules, which trace $\sim$1.7~bar. For WISE~1049A, the maximum deviation varies between 1.24\% and 1.58\% for the molecular features tracing $\sim$3~bar, and less variability, between 0.81\% and 1.25\% for the molecular features tracing up, $\sim$1.7~bar. Considering the uncertainties around $\sim$0.06-0.12\%, calculated as the standard deviation of the maximum and minimum peaks of the light curves, and thanks to the accuracy of the JWST data (explained in detail in Section \ref{amplitudes} and listed in the Table \ref{table:limits_molecules}), we can consider relevant the variation in the maximum deviation for light curves that trace similar pressure levels. For WISE~1049B, we observe the same differences in the maximum deviation for the $\mathrm{CH_{4}}$-3, $\mathrm{CH_{4}}$-5, and $\mathrm{H_{2}O}$-1 molecules, which range between 4.85\% and 6.3\%. However, the differences in the maximum deviation for the molecular features that trace $\sim$1.7~bar do not change significantly considering the uncertainties, ranging from 5.19\% to 5.09\% when considering the maximum deviation calculated with the fits. Therefore, it is important to note that in some cases, different molecular features at the same pressure may be affected differently by physical processes, generating variability in that region of the atmosphere.

    \item For WISE~1049A, the maximum deviation measured for the CO and CH$_4$ molecular features does not decrease towards the top of the atmosphere, as expected if the only source of variability were silicate clouds, because the longer wavelengths that probe the upper atmosphere would not be affected by the clouds in the deep zone. This is not the case for the H$_2$O molecular features, for which their maximum deviation decreases toward the top of the atmosphere. Instead, for the CO and CH$_4$ molecular features, we observe a minimum in the variability at the middle pressures of the atmosphere (P $\sim$~1.6 bar). If the only source of variability were clouds, which are found at the bottom of the atmosphere, we would expect to observe a monotonically decreasing maximum deviation for all the light curves as they approach the top of the atmosphere. This suggests that an additional physical mechanism is needed to explain this variability in the CO and CH$_4$ molecular features.

\end{enumerate}

As previously mentioned in Section \ref{sec:intro}, one proposed mechanism in the literature that might explain the behavior of the maximum deviation observed for WISE~1049A light curves is thermochemical instabilities \citep{tremblin2015fingering, tremblin2016cloudless}.  Thermochemical instabilities may occur when heat is transported in the atmosphere of an object, altering the concentration of molecules in the atmosphere and thereby changing the chemical equilibrium of the reactions that are occurring. For the case of L/T transition brown dwarfs, the most important chemical reaction altered would be equation \ref{eq:reaction}, involving CO and $\mathrm{CH_{4}}$; however, it will also apply to other molecules that we are not observing in this wavelength coverage. Thus, thermochemical instabilities might influence the variability for those molecular bands, as proposed by \citet{tremblin2020rotational}, using 1D models. 

On the other hand, these chemical instabilities could also arise from feedback due to the presence of clouds in the atmosphere, as shown in several General Circulation Models (GCMs). \cite{lee2023dynamically, lee2024dynamically} show that vertical convection and storms locally alter the abundances of key gases (such as CO and CH$_4$), since both cloud particles and chemical species are driven both dynamically and by temperature. In their models, vertical mixing and local cooling levels induce storm-induced cloud convection, which increases the chemical imbalance in the most cloudy regions. They also conclude that cloud opacity modifies local temperatures, which in turn affects the chemical response, thereby reinforcing spatial heterogeneities in species abundance.

Based on the correlations between the light curves of different molecular features discussed in Section \ref{sec:lc-correlations}, where we calculated the Kendall coefficients for all light-curve combinations, we find a strong correlation between the CH$_4$ and H$_2$O features. This behavior is expected because both molecules lie on the same side of Equation \ref{eq:reaction}, leading them to vary in tandem. In contrast, when we compare the CO light curves with those of CH$_4$ and H$_2$O for both WISE 1049A and WISE 1049B, we find no significant correlation—likely a consequence of the chemical reaction in Equation \ref{eq:reaction}. Furthermore, although the same behavior is observed for both objects, the correlations for WISE 1049A are weaker in all cases. Therefore, chemical disequilibrium would be affecting WISE 1049A to a greater extent than WISE 1049B, which is what we are observing in our results. This may represent one of the first tentative observational indications of brown dwarf variability due to disequilibrium chemistry, as the affected molecules are CH$_4$ and CO, which are involved in thermochemical reactions, confirming the expectation mentioned in \citealt{biller2024jwst}, \citealt{mccarthy2024jwst}, and \citealt{chen2025jwst}. We do not rule out the possibility that other mechanisms could influence them to a lesser extent.

On the other hand, for WISE~1049B, the effect of thermochemical instability is not as clear as in the case of A. The variability observed for WISE~1049B could be explained only by the existence of clouds in its atmosphere. 
\vspace{5mm}

\subsection{Artistic representation of structures in WISE 1049AB}

\begin{figure}
    \centering
    \includegraphics[width=1\linewidth]{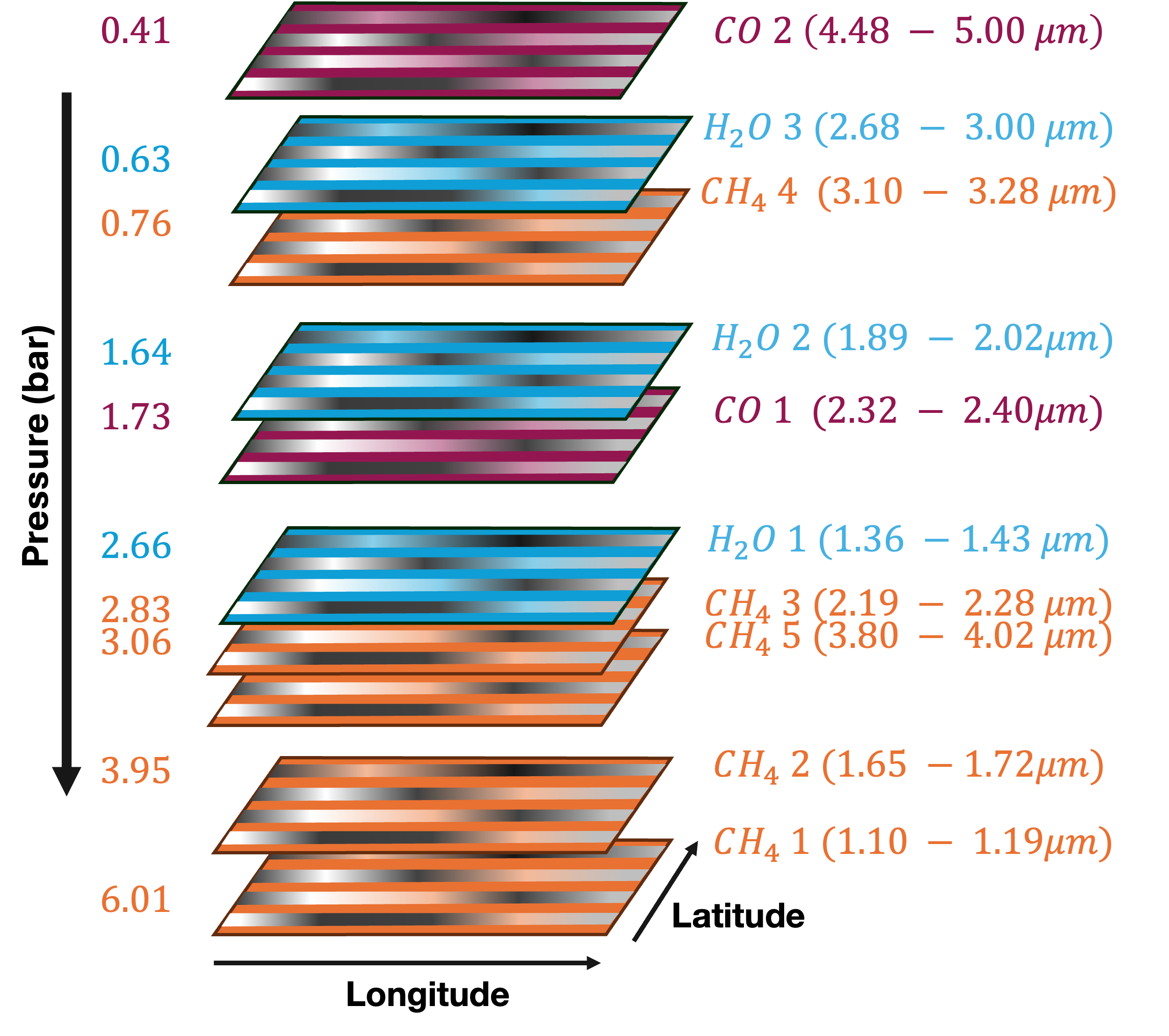}
    \caption{Artistic representation of the atmospheric layers of WISE 1049A, traced by each molecular feature. Each degraded band represents one planetary wave in each layer, where contrasting the lighter and darker regions with the rotation would generate the variability.}
    \label{fig:pressure-structures_A}
\end{figure}

\begin{figure}
    \centering
    \includegraphics[width=1\linewidth]{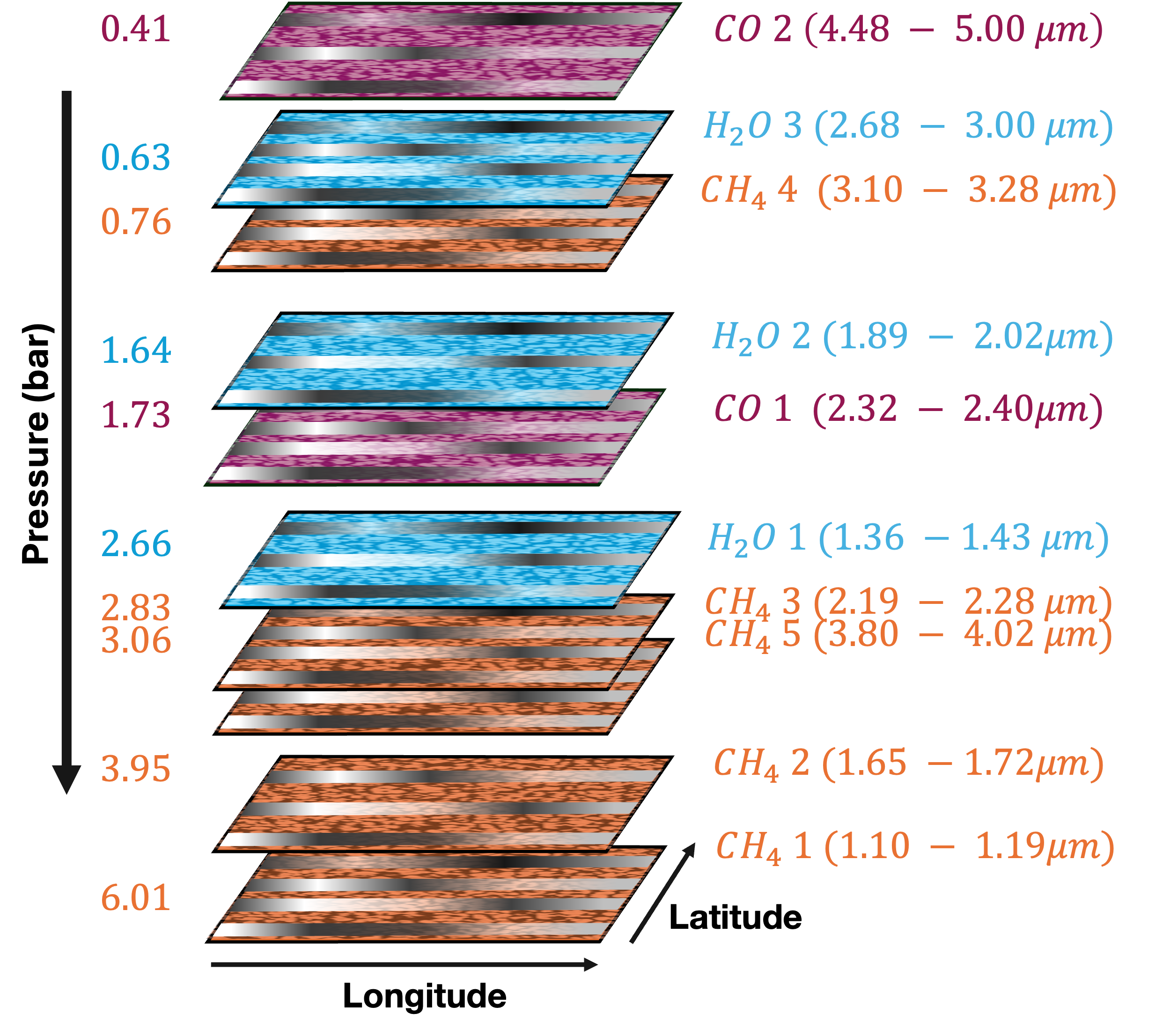}
    \caption{Artistic representation of the atmospheric layers of WISE 1049B, traced by each molecular feature. Each degraded band represents one planetary wave in each layer, where contrasting the lighter and darker regions with the rotation would generate variability. The background structures represent the possible cloud spots or vortices that cannot be explained solely by planetary waves.}
    \label{fig:structures_B}
\end{figure}

We used the best sinusoidal fits and their residuals found in Section \ref{structures_map} (see Figures \ref{fig:lc-A} and \ref{fig:lc-B})  for each molecular feature, and the pressure levels traced by each molecular band given by the contribution function in Figure \ref{fig:cont_func}, to reproduce a 3D illustration or artistic representation of the expected clouds structures at each pressure level or depth in the atmosphere of WISE~1049AB (see Figure \ref{fig:pressure-structures_A} and \ref{fig:structures_B} respectively).

The best sinusoidal fits obtained in Section \ref{structures_map} represent a planetary-scale wave at a specific pressure level, for each molecular feature, meaning that waves can organize atmospheric circulation into banded structures and/or localized spots. The number of sinusoidal waves that best match each of the light curves presented in Section \ref{structures_map} represents the number of planetary-scale waves that might be present in the atmosphere of the object at each pressure level. If the residuals remaining after we subtract those sinusoidal fits from the light curve have a structure or have remaining periodic patterns, they might suggest the existence of cloud spots or vortices that are not reproduced by the sinusoidal fits. We attributed vortex-like structures to the GCM results of \citealt{mukherjee2021modeling}. Their Fig. 4 suggests that zonal jets and vortices can form in these atmospheres. The jets affect the light curves on larger time scales, and the vortices are the ‘leftover’ signals. However, it is important to note that since the light curves presented here only cover one rotational period for each object, we are unable to provide the exact location of these features in terms of longitude and latitude, this is why the representations shown here in Figures  \ref{fig:pressure-structures_A} and \ref{fig:structures_B} are considered illustrative representations. In addition, since this system appears to evolve rapidly \citep{fuda2024latitude}, the map for one rotation may differ from that for the next.

Figure \ref{fig:pressure-structures_A} provides a qualitative representation of the layered atmosphere of WISE 1049A, based on the pressure levels indicated by the contribution function in Table \ref{table:limits_molecules}. The atmosphere of WISE 1049A shows the same atmospheric map at all depths, since all the light curves tracing those pressure levels are better matched by four sinusoidal functions, as described in Section \ref{structures_map} (Figure \ref{fig:lc-A}, right). However, while four sinusoids do fit all wavelengths for WISE 1049A, the sinusoids (and light curve shapes) are quite different, causing notable ``phase shifts", as noted by \citet{biller2024jwst} (see their Fig. 8); however, we do not consider the quantitative values of the shifts. In addition, given that the residuals between the fit and the data are flat, we infer that WISE~1049A might not have spots of clouds in its atmosphere (see Figure \ref{fig:lc-A}). The colors used in Figure \ref{fig:pressure-structures_A} of each layer are not real, and they are color-coded to match the corresponding molecular band light curve we used to obtain the artistic representation of a given layer.

The light curves of WISE 1049 B were reproduced by three or four sinusoidal waves, depending on the wavelength range (or depth). Independently of the number of waves needed, the residuals remaining after subtracting those sinusoidal waves have a structure (see Figure \ref{fig:lc-B}), suggesting that more complex cloud structures not reproduced by the sinusoidal fits might also be present in the atmosphere of WISE~1049B, in contrast to A. We present a qualitative representation of the layered atmosphere WISE~1049AB in Figure \ref{fig:structures_B}, where we can observe that the cloud structures present at different depths in the atmosphere are different. 
As mentioned above, since the light curves presented here only cover one rotational period for each object, we are unable to obtain an exact map of the object; thus, what is shown in Figure \ref{fig:pressure-structures_A} is just an artistic representation. Since not all light curves exhibit the peak of the residuals at the same time, different cloud spots or vortices are likely present in each atmospheric layer. For example, the peak in the residuals of the CH$_4$-4 molecular band with ranges of 3.10 -- 3.28 $\mu$m in the spectra is located around 2.5 hr, while for most other CH$_4$ molecular bands, this peak is found around 1.1 hr, which means that the residuals show a phase shift.

\section{Discussion}\label{sec:discussion}

\subsection{Comparison with previous studies of WISE 1049AB}

We found consistent rotational periods for WISE~1049~AB with those derived by other studies (6--8~hr for WISE 1049A and $\sim$5~hr for WISE 1049B, \citealt{gillon2013fast}, \citealt{burgasser2014monitoring}, \citealt{buenzli2014brown}, \citealt{Apai2021}, \citealt{fuda2024latitude}), not only from a light curve in a particular photometric band, but even using the light curves of different molecular bands at different depths. Furthermore, our results are consistent with the most recent results of \citet{biller2024jwst}, and \citet{chen2025jwst}, using a more extended time baseline, since they combine the MIRI and NIRSpec observations of both epochs of program GO~2965.

Our calculations of the maximum deviation for the different molecular bands agree with the studies reported in the literature, where a higher maximum deviation is found for WISE 1049B than for WISE 1049A. We found a maximum deviation of 7.00\% for WISE 1049B, for the CH$_4$-1 molecular feature in the range of 1.10--1.19~$\mu$m, while the maximum deviation of  WISE 1049A was 2.66\% for the CO-2 molecular feature in the range of 4.48--5.00~$\mu$m.

\citet{gillon2013fast} found $\sim11$~\% variability for the system observing at 910~nm.  \citet{burgasser2014monitoring} found a 13.5 \% variability amplitude for the WISE 1049B object, and no variability in A in the 0.8–2.4~$\mu$m wavelength range. In addition, \citet{buenzli2015cloud} found a variability of 7--11\% for WISE 1049B, and a variability of 0.4\%  for WISE 1049A. 
In contrast to these studies, we found a maximum variability of 2.66\% for WISE 1049A, which may be attributed to the higher sensitivity of JWST compared to previous space-based and ground-based telescopes or to the evolution of cloud structures in the object's atmosphere over time.

We found a similar number of fits, periods, and variability amplitudes to those reported by \cite{apai2017zones} and \cite{fuda2024latitude}, using TESS monitoring (600–1000~nm) during $\sim$100 rotations of WISE~1049AB. \cite{fuda2024latitude} fitted several sinusoidal functions to those data, finding between three and four sinusoidal waves in different light curves obtained at different epochs and with 30 and 80 hr long light curves, respectively. We found a consistent rotational period for all of the molecular bands. The best-fit parameters found by \citealt{fuda2024latitude} are comparable with the parameters found for our best fits for WISE 1049B, where we obtained periods around 5.0 and 2.5~hr, and with the maximum deviations around 3.43 -- 7.00 \%.

Using HST/WFC3 photometry, \citet{karalidi2016maps} retrieved a map of WISE~1049AB with three or four spots of clouds at the top of the atmosphere of WISE 1049A and B, using \texttt{Aeolus}, with a surface coverage of 19--32~\% (depending on an assumed rotational period of 5 hr or 8~hr).  In a recent study by \citet{chen2024global}, possible cloud spots on the surface of WISE 1049B were identified, which is consistent with our results of bands and cloud spots for different molecular bands at varying depths. 

\subsection{Variability Mechanisms}

Both clouds and thermochemical instabilities appear to be necessary to explain the observed variability in WISE~1049A. As explained in Section \ref{Sec:physicalmec}, the increased maximum deviations of the light curves that trace the upper layers of the atmosphere of WISE~1049A ($\mathrm{H_{2}O}$-3 and CO-2) cannot be explained only by the existence of silicate clouds, since those are found primarily on lower layers of the atmosphere (see Figure \ref{fig:mechanism-A}, right panel). The existence of thermochemical instabilities in WISE~1049A might explain the increased variability at the top of its atmosphere for the $\mathrm{H_{2}O}$-3 and CO-2 molecules. However, the variability measured in the light curves that trace the bottom of the atmosphere ($\mathrm{CH_{4}}$-1, $\mathrm{CH_{4}}$-2, $\mathrm{CH_{4}}$-5 and $\mathrm{CH_{4}}$-3) is most likely due to silicate clouds.

As explained in Section \ref{Sec:physicalmec}, for WISE~1049B, the maximum deviation decreases almost monotonically from the bottom to the top of the atmosphere. As for WISE~1049A, the silicate clouds are situated at the bottom of the atmosphere, and they are able to explain the variability observed for all of the light curves for this object.

\cite{millar2020detection} measured the polarization signal in the $H$-band of WISE~1049AB using NaCo at the Very Large Telescope, finding a polarization signal for WISE 1049A of $p_A  = 0.031~\%~\pm 0.004\%$, and for WISE 1049B to be $p_B  = 0.010~\%~\pm~0.004~\%$. The fact that both objects exhibit a significant polarization signal further supports the likely presence of clouds in their atmospheres. However, WISE1049A shows a higher polarization signal than B. This suggests that A might be potentially more cloudy than B, which is expected given the late-L spectral type of A. Alternatively, the clouds of WISE~1049~A could be more inhomogeneous, which would explain its higher polarization degree.

The results broadly agree with the predictions of GCMs adapted for WISE 1049B presented by \citet{chen2025jwst}. The GCM simulations predict that the variability seen in the near-infrared (1.0 -- 2.5 $\mu$m) that originates from deeper in the atmosphere is due to the effect of patchy clouds rotating into and out of view, which is directly correlated with the increasing opacity of the selected molecular bands. This variability is dynamically induced by thermal feedback from cloud formation and variations in cloud vertical extent and opacity across the globe. Overall, the GCM simulations qualitatively agree with the results presented in this study: the molecular bands selected from that region of the spectrum trace lower depths, where most clouds are, and show the highest maximum deviation.

We also compared our results with the hot spot chemical predictions \citep{morley2014spectra}, as presented in Figure 15 of \citealt{chen2025jwst} for WISE 1049B, and with Figure 7 of \citet{mccarthy2024jwst} for SIMP 0136, a T2.0 object with a similar temperature to WISE 1049B. In both cases, we can observe that the fluctuations of temperature affect the max/min ratios (presented in Figure \ref{fig:spectrum_A}), mainly between 2.5 and 3.5 $\mu$m, i.e., near the middle part of the atmosphere, according to our contribution functions. It is in this wavelength range that we observe the effects attributed to temperature fluctuations and their impact on disequilibrium chemistry, which affect the CO and CH$_4$ molecular features. These features do not follow a trend, unlike those affected mainly by clouds deep in the atmosphere. 

\citet{biller2024jwst}, \citet{mccarthy2024jwst}, and \citet{chen2025jwst} used K-means clustering to find groups of similar shapes of light curves of different wavelength ranges. Additionally, when they compare the pressure levels traced by each cluster in the atmosphere, they can conclude that the distinct shapes of the light curves in each cluster suggest different mechanisms generate the variability responsible for each cluster. Each cluster traces a specific pressure level in the atmosphere.

\cite{biller2024jwst} and \cite{mccarthy2024jwst} suggested that thermochemical instabilities could affect each molecular band differently. In our analysis, we confirm this by comparing the shapes and the difference in the maximum deviation of two different molecular features associated with other molecules, for example, H$_2$O-3 and CH$_4$-4. In Figure \ref{fig:mechanism-A} and \ref{fig:mechanism-B}, we observed that CH$_4$-4 in the range 3.10~--~3.28 $\mu$m and H$_2$O-3 in the range 2.68 -- 3.0 $\mu$m trace similar pressure $\sim$0.7~bar for WISE 1049A and B, but they have different amplitudes and their shapes change a little, and it may be due to the CH$_4$-4 being more affected by thermochemical instabilities.

In addition, it is essential to consider the influence of the rotation and inclination (viewing angle) on the variability measured for brown dwarfs. WISE 1049B is a faster rotator ($\sim$5~hr) than WISE 1049A ($\sim$7~hr). \citet{tan2021atmospheric-a, tan2021atmospheric-b} show that radiative feedback from clouds can generate vigorous and patchy weather patterns, the nature of which depends mainly on rotation and the angle of observation. Due to its faster rotation, this implies that in WISE 1049B, the fragmented clouds could generate more changeable structures, resulting in greater maximum deviation, while in WISE 1049A, which rotates slightly slower, it might develop broader and more persistent storms, manifesting a more periodic modulation in its light. Also, these two objects have different viewing angles between them, where A is more pole on (within $62^\circ$), while B is more equator on ($\sim 90^\circ$, \citealt{Apai2021}). \citet{Vos2017} concluded that equator-on brown dwarfs are redder and they show higher maximum deviation, which is in agreement with what we observe for WISE 1049AB.  We observe the same conclusions that \citet{tan2021atmospheric-a} predicts, greater maximum deviation for objects observed equatorially than for objects observed from above, which is also corroborated by observational studies such as \citet{vos2020spitzer} and \citet{Suarez2023}.

\section{Conclusions}\label{Sec:conclusions}

We monitored the WISE 1049AB binary system using JWST/NIRSpec for $\sim$8 hrs \citep{biller2024jwst}. We performed time-resolved spectroscopy of both components in the 0.6 -- 5.3 $\mu$m wavelength range to characterize their variability. In particular, we characterized the variability of five molecular absorption bands of CH$_4$, three of H$_2$O, and two of CO (the wavelength ranges are summarized in Table \ref{table:limits_molecules}). We organized the light curves in terms of atmospheric depth, based on the contribution function of the system derived using atmospheric models. This allows us to trace how the variability changes at different depths, probably due to various physical mechanisms at play in the WISE~1049~AB atmospheres.

\begin{enumerate}

\item We measured the rotational periods and amplitudes for both brown dwarfs and found results consistent with previous studies: $\sim$7~hr for WISE~1049A and $\sim$5~hr for WISE~1049B.  In addition, we found WISE~1049A less variable than WISE~1049B, as found by previous studies. WISE~1049A shows maximum deviation ranging from 0.70\% to 2.66\%, while WISE~1049B showed higher variability, with amplitudes between 3.43\% and 7.00\%.

\item Measuring the maximum deviation of the different molecular bands that trace the atmosphere of WISE~1049~AB, we concluded that only clouds, mostly expected at the bottom of the atmosphere of WISE~1049AB, are not able to explain the variability measured in the mid- and upper atmosphere of WISE~1049~A. The upper atmosphere of WISE~1049A exhibits variability similar to that measured in the deep atmosphere. Thermochemical instabilities might explain the increased variability of amplitudes measured in the upper atmosphere of WISE~1049A traced by the CO-2  molecular band.

\item The variability observed for WISE~1049B can be explained only with the existence of silicate clouds. We cannot rule out that other mechanisms may be involved, but they would not contribute as significantly.

\item We obtained the first three-dimensional artistic representation of WISE~1049AB using sinusoidal functions to fit the different light curves of both components in the system, corresponding to various molecular bands listed in Table \ref{table:limits_molecules} and their residuals. WISE~1049A has four bands at all depths and exhibits flat residuals, indicating that its variability may be explained by the bands alone. However, WISE~1049B most likely has three or four bands, depending on the atmospheric depth. The residuals of WISE~1049B suggest the presence of more complex structures, such as vortices and cloud spots, that cannot be explained by sinusoids alone. The light curves presented here do not cover the 2--3 rotational periods needed to constrain the exact longitudinal and latitudinal location of these atmospheric features.


\end{enumerate}


\section{Acknowledgements}
NOG acknowledges the support from the Arthur Davidsen Graduate Student Research Fellowship provided by the Space Telescope Science Institute. BB and BJS acknowledge funding by the UK Science and Technology Facilities Council (STFC) grant no. ST/V000594/1. AMM acknowledges support from the National Science Foundation Graduate Research Fellowship Program under Grant No. DGE-1840990. JMV acknowledges support from a Royal Society - Research Ireland University Research Fellowship (URF/1/221932).

The JWST data presented in this article were obtained from the Mikulski Archive for Space Telescopes (MAST) at the Space Telescope Science Institute. The specific observations analyzed can be accessed via \dataset[doi: 10.17909/kzxq-e127]{https://doi.org/10.17909/kzxq-e127}.

This work made use of Astropy:\footnote{http://www.astropy.org} a community-developed core Python package and an ecosystem of tools and resources for astronomy \citep{astropy:2013, astropy:2018, astropy:2022}.

\bibliography{sample631}{}
\bibliographystyle{aasjournal}

\appendix
\section{Parameters and fits}\label{fits}

\begin{deluxetable*}{ccccccccc}[h]
\centering
\tablecaption{Parameters and Fit Results for the multi-sine models for the A object. \label{Table:ParametersFitA}}
\tablehead{
\colhead{Method} & \colhead{Molecular} & \colhead{N$^\circ$} & \colhead{Amplitude} & \colhead{Period} & \colhead{Phase} & & \colhead{Value} & \\
\colhead{} & \colhead{band} & \colhead{sin} & \colhead{(\%)} & \colhead{(hr)} & \colhead{(radian)} & \colhead{simpler} & \colhead{best} & \colhead{complex}
}
\startdata
$\Delta AIC$    &          &            &                                 &                              &                       & 449.00      & 0.00    & 47.65  \\
$\Delta BIC$   & CH$_4$-1    & 4          & 0.27, 0.30, 0.06, 2.11     & 3.08, 3.14,   & -0.01, -0.01, 0.01, -0.01   & 411.92 & 0.00 & 29.11 \\
\hline
$\Delta AIC$    &          &            &                                 &                              &                        & 1574.01  & 0.00  & 721.79  \\
$\Delta BIC$   & CH$_4$-2    & 4          & 1.41, 0.11, 0.11, 0.25 & 3.39, 3.75,   & -0.01, 0.38, 0.39, 0.04 & 1536.95 & 0.00 & 1262.08 \\
\hline
$\Delta AIC$    &          &            &                                 &                              &               & 8503.28 & 0.00 & 4.37   \\
$\Delta BIC$   & CH$_4$-3    & 4          & 1.14, 3.44, 1.37, 0.04   & 3.30, 2.47,  & -0.01, -0.22, -0.03, -0.02  & 8501.37 & 0.00 & 14.17 \\
\hline
$\Delta AIC$    &          &            &                                 &                       &       & 432.01 & 0.00 & 345.63 \\
$\Delta BIC$   & CH$_4$-4    & 4          & 1.61, 0.72, 0.08, 0.76  & 3.32, 4.72,   & -0.35, 0.04, 0.02, -0.01   & 394.95 & 0.00 & 327.09 \\
\hline
$\Delta AIC$    &          &            &                                 &                              &      & 671.03  & 0.00 & 9668.39  \\
$\Delta BIC$   & CH$_4$-5    & 4          & 0.05, 0.06, 2.01, 4.34    &14.02, 3.98,  & -0.01, -0.01, -0.01, 0.02   & 652.50 & 0.00 & 9670.92 \\
\hline
$\Delta AIC$    &          &            &                                 &                              &          & 7989.65     & 0.00  & 438.49  \\
$\Delta BIC$   & H$_2$O-1    & 4          & 1.90, 1.53, 1.53, 0.04 & 6.02, 3.62,  & -0.11, -0.01, 0.01, 0.01 & 787.21 & 0.00 & 419.96 \\
\hline
$\Delta AIC$    &          &            &                                 &                              &                         & 920.92   & 0.00  & 7.55    \\
$\Delta BIC$   & H$_2$O-2    & 4          & 3.44, 1.37, 1.21, 0.07   & 3.45, 1.86,   & -0.10, -0.01, 0.01, 0.01   & 894.84 & 0.00 & 10.98 \\
\hline
$\Delta AIC$    &          &            &                                 &                              &      & 1017.19 & 0.00 & 35.38  \\
$\Delta BIC$   & H$_2$O-3    & 4          & 3.54, 1.34, 1.26, 0.06   & 3.46, 7.91,   & -0.11, -0.01, 0.01, 0.01   & 1029.07 & 0.00 & 16.85  \\
\hline
$\Delta AIC$    &          &            &                                 &                              &     & 815.24    & 0.00 & 12.90  \\
$\Delta BIC$   & CO-1     & 4          & 0.65, 2.03, 0.08, 0.08  & 3.18, 6.96,   & -0.02, 0.02, -0.02, 0.11 & 815.71 & 0.00 & 31.43 \\
\hline
$\Delta AIC$    &          &            &                                 &                              &     & 793.91  & 0.00 & 912.35 \\
$\Delta BIC$   & CO-2     & 4          & 0.31, 0.76, 4.75, 0.06   & 7.58, 2.08,    & 0.05, -0.01, 0.03, 0.01   & 777.96 & 0.00 & 912.96 \\
\enddata
\tablecomments{We report $\Delta$AIC and $\Delta$BIC (differences relative to the minimum AIC/BIC for each band). The best model has $\Delta = 0$. Larger $\Delta$ values indicate weaker support; $\Delta > 10$ is considered strong evidence against the model. Definitions follow the Gaussian log-likelihood formulation (see Section~4.3). Periods longer than 8 hours should be interpreted with caution.}
\end{deluxetable*}

\begin{deluxetable*}{ccccccccc}[h]
\centering
\tablecaption{Parameters and Fit Results for the multi-sine models for the B object.  \label{Table:ParametersFitB}}
\tablehead{
\colhead{Method} & \colhead{Molecular} & \colhead{N$^\circ$} & \colhead{Amplitude} & \colhead{Period} & \colhead{Phase} & & \colhead{Value} & \\
\colhead{} & \colhead{band} & \colhead{sin} & \colhead{(\%)} & \colhead{(hr)} & \colhead{(radian)} & \colhead{simpler} & \colhead{best} & \colhead{complex}
}
\startdata
$\Delta AIC$                       &          &                          &                             &                             &                      &   3240.48   & 0.00 & 816.90 \\
$\Delta BIC$                       & CH$_4$-1 & 3                        & 0.02, 5.01, 1.00            & 1.46, 5.83,                 & -1.43, -0.67, 1.49   &   3237.42   & 0.00 & 798.37 \\
\hline
$\Delta AIC$                       &          &                          &                             &                             &                      &   3408.80   & 0.00 & 3946.96 \\
$\Delta BIC$                       & CH$_4$-2 & 3                        & 1.35, 0.03, 1.35            & 3.90, 2.87,                 & 3.78, 1.97, 0.62      &   3403.19   & 0.00 & 3854.29 \\
\hline
$\Delta AIC$                       &          &                          &                             &                             &                      &   9395.63   & 0.00 & 5.28 \\
$\Delta BIC$                       & CH$_4$-3 & 3                        & 0.02, 2.70, 0.013           & 1.34, 5.40,                 & -7.66, -6.90, 12.19   &   9393.10   & 0.00 & 23.81 \\
\hline
$\Delta AIC$                       &          &                          &                             &                             &                      &   9802.13   & 0.00 & 2952.57 \\
$\Delta BIC$                       & CH$_4$-4 & 4                        & 1.59, 0.02, 1.58, 2.21      & 4.40, 2.95,                 & -0.93, -3.86, 2.20, 2.67 & 9799.08 & 0.00 & 2934.05 \\
\hline
$\Delta AIC$                       &          &                          &                             &                             &                      &   9504.82   & 0.00 & 584.29 \\
$\Delta BIC$                       & CH$_4$-5 & 4                        & 0.02, 1.96, 1.96, 2.54      & 2.70, 5.31,                 & -4.69, 1.84, 1.84, 2.73 & 9500.76  & 0.00 & 565.76 \\
\hline
$\Delta AIC$                       &          &                          &                             &                             &                      &   95.19     & 0.00 & 863.07 \\
$\Delta BIC$                       & H$_2$O-1 & 4                        & 0.03, 1.29, 1.27, 5.03      & 5.10, 2.66                  & -1.16, -7.72, -1.44, -3.34 & 92.11  & 0.00 & 844.52 \\
\hline
$\Delta AIC$                       &          &                          &                             &                             &                      &   2662.84   & 0.00 & 15.03 \\
$\Delta BIC$                       & H$_2$O-2 & 3                        & 0.02, 0.01, 2.29            & 2.77, 5.29                  & -4.66, -0.76, 2.69     & 2662.91   & 0.00 & 20.60 \\
\hline
$\Delta AIC$                       &          &                          &                             &                             &                      &   9682.44   & 0.00 & 9746.33 \\
$\Delta BIC$                       & H$_2$O-3 & 4                        & 0.01, 0.01, 0.01, 1.17      & 2.48, 1.96,                 & -1.35, -7.27, -6.25, 1.02 & 9680.91  & 0.00 & 9748.86 \\
\hline
$\Delta AIC$                       &          &                          &                             &                             &                      &   116.47    & 0.00 & 70.16 \\
$\Delta BIC$                       & CO-1     & 4                        & 0.02, 0.01, 0.01, 1.22      & 2.79, 5.76                  & -6.57, -0.03, -0.99, 0.97 & 112.41   & 0.00 & 51.64 \\
\hline
$\Delta AIC$                       &          &                          &                             &                             &                      &   1132.25   & 0.00 & 578.07 \\
$\Delta BIC$                       & CO-2     & 3                        & 0.05, 2.60, 0.01            & 1.59, 5.51                  & -2.37, 5.89, -0.10     & 1128.20   & 0.00 & 559.55 \\
\enddata
\tablecomments{We report $\Delta$AIC and $\Delta$BIC (relative to the minimum value for each criterion and molecular band). By definition, the best model has $\Delta=0$. Larger $\Delta$ indicates less support; $\Delta>10$ is generally strong evidence against the model. Definitions use the full Gaussian log-likelihood (see Section~4.3). }
\end{deluxetable*}

\clearpage

\section{Continuum and absorption features}\label{AppendixB}

We measured the variability of the continuum around the molecular bands mentioned in this article to test whether the variability remains the same within the molecular band and in the continuum surrounding it. Thus, we also measure the light curves for these regions of the pseudo continuum around each molecular absorption. In Table \ref{continum-abs}, we show the selected regions for each molecular band's continuum and the relative variability value we measured in each of them, following the method described in Section \ref{amplitudes}.

We confirm that: (1) All the light curves of the molecular band selected in this article have the same shape as their adjacent continuum, since they probe similar pressures, that means, they are affected by the same physical mechanisms. However, the maximum deviation could change significantly in the molecular bands with respect to their adjacent continuum, depending on the wavelengths. For the molecular features around 1.5-2.5 $\mu$m, as H$_2$O-2 continuum, they trace the J and H band peaks of the spectrum, which are intrinsically more variable since they trace the bottom of the atmosphere where the silicate clouds are expected, showing that the values at the continuum and absorption are higher enough considering their errors. But for the cases of longer wavelengths ($>$3.1 $\mu$m), the continuum light curves have lower maximum deviations. The CO-1 ‘continuum’ overlaps with part of the absorption of the H$_2$O-3 and CH$_4$-4 molecular bands. In this particular case, it is more difficult to identify the pseudo-continuum, and the shape of the light curves changes. (2) The shape of the light curves in the relative variability of the continuum is the same as for the absorption, but at a lower amplitude. 

\begin{deluxetable*}{ccccccc}[h]
\tablecaption{Continuum and absorption wavelength ranges with their respective values of the maximum deviation.}\label{continum-abs}
\tablehead{
\colhead{  Molecular  } & \colhead{Absorption} &  \colhead{Continuum} & \colhead{Max. dev. Cont.} &\colhead{Max. dev. Abs.} & \colhead{Max. dev. Cont.} & \colhead{Max. dev. Abs.}  \\
\colhead{band} & \colhead{range ($\mu$m)} & \colhead{range ($\mu$m)} & \colhead{WISE 1049A (\%)} &\colhead{WISE 1049A (\%)} & \colhead{WISE 1049B (\%)} &  \colhead{WISE 1049B (\%)}
}
\startdata
CH$_4$ 1 & 1.10 -- 1.19 & 0.90 -- 1.10 \& 1.19 -- 1.30 & 2.01$\pm$ 0.12 & 2.06 $\pm$ 0.11 & 6.24$\pm$ 0.08  & 7.00$\pm$ 0.06 \\
H$_2$O 1 & 1.36 -- 1.43 & 1.19 - 1.30 \& 1.51 -- 1.58  & 1.61$\pm$ 0.12 & 1.58$\pm$ 0.14 & 6.07$\pm$ 0.09  & 6.30$\pm$ 0.08 \\
CH$_4$ 2 & 1.65 -- 1.72 & 1.51 -- 1.60 \& 1.75 -- 1.60 & 0.47$\pm$ 0.07 & 1.35 $\pm$ 0.05& 6.59$\pm$ 0.06  & 6.79$\pm$ 0.04 \\
H$_2$O 2 & 1.89 -- 2.02 & 1.57 -- 1.75 \& 2.05 -- 2.22 & 1.37$\pm$ 0.09 & 1.25$\pm$ 0.09 & 5.84$\pm$ 0.05  & 5.75$\pm$ 0.04 \\
CH$_4$ 3 & 2.19 -- 2.28 & 2.12 -- 2.19 \& 2.28 -- 2.39 & 0.48$\pm$ 0.11 & 1.30 $\pm$ 0.10 & 5.15$\pm$ 0.06  & 5.72$\pm$ 0.05 \\
CO 1     & 2.32 -- 2.40 & 2.20 -- 2.30 \& 2.50 -- 2.60 & 1.47$\pm$ 0.10 & 0.81 $\pm$ 0.10 & 5.51$\pm$ 0.07  & 5.21$\pm$ 0.06 \\
H$_2$O 3 & 2.68 -- 3.00 & 2.40 -- 2.50 \& 3.00 -- 3.10 & 0.62$\pm$ 0.06 & 1.09 $\pm$ 0.05 & 3.77$\pm$ 0.08  & 3.49$\pm$ 0.06 \\
CH$_4$ 4 & 3.10 -- 3.28 & 2.90 -- 3.10 \& 3.50 -- 3.70 & 0.30$\pm$ 0.06 & 0.70 $\pm$ 0.05 & 3.82$\pm$ 0.06  & 4.40$\pm$0.08 \\
CH$_4$ 5 & 3.80 -- 4.02 & 3.40 -- 3.60 \& 4.10 -- 4.23 & 0.36$\pm$ 0.06 & 1.24$\pm$ 0.05 & 3.79$\pm$ 0.05  & 4.85$\pm$ 0.05 \\
CO 2     & 4.48 -- 5.00 & 4.10 -- 4.30 \& 5.00 -- 5.30 & 2.14$\pm$ 0.07 & 2.66 $\pm$ 0.06 & 1.65$\pm$ 0.07  & 3.43$\pm$ 0.06
\enddata
\tablecomments{The uncertainties of the relative amplitudes (\%) are associated with the standard deviation of the values around the maximum and minimum peak of the observational data}.
\end{deluxetable*}

\section{Contribution Function details}\label{AppendixC}

We searched for the best fit to the spectrum of each object using a custom-made grid of models created using \texttt{picaso} \citep{Mukherjee_2023} and input from Sonora Bobcat  \citep{marley2021sonora} and Diamondback models \citep{Morley_2024}. We used $\log K_\mathrm{zz}$ ranging from $9$ to $12$ and $f_\mathrm{sed}=1, 2, 3, 4$ and $8$. Our best-fit models found a temperature of $T_{eff}$=1500~K and surface gravity of log~$g$=4.0 for WISE 1049A, and $T_{eff}$=1400~K and log~$g$=3.5 for WISE 1049B, consistent with recent results presented in \citet{ishikawa20251} and \citet{de2025eso}. However, using the ExoRem model grid \citealt{charnay2018self}, we found the best fit with $T_{eff}$=1300~K and gravity of log~$g$=4.5 for WISE 1049A, and $T_{eff}$=1250~K and log$g$=5.0 for WISE 1049B that is more consistent with the results of \citet{biller2024jwst} where they found an effective temperature for both objects of 1200--1300~K and a log~g of 5.0 using evolutionary models. 

\begin{figure*}
    \centering
    \includegraphics[width=0.9\linewidth]{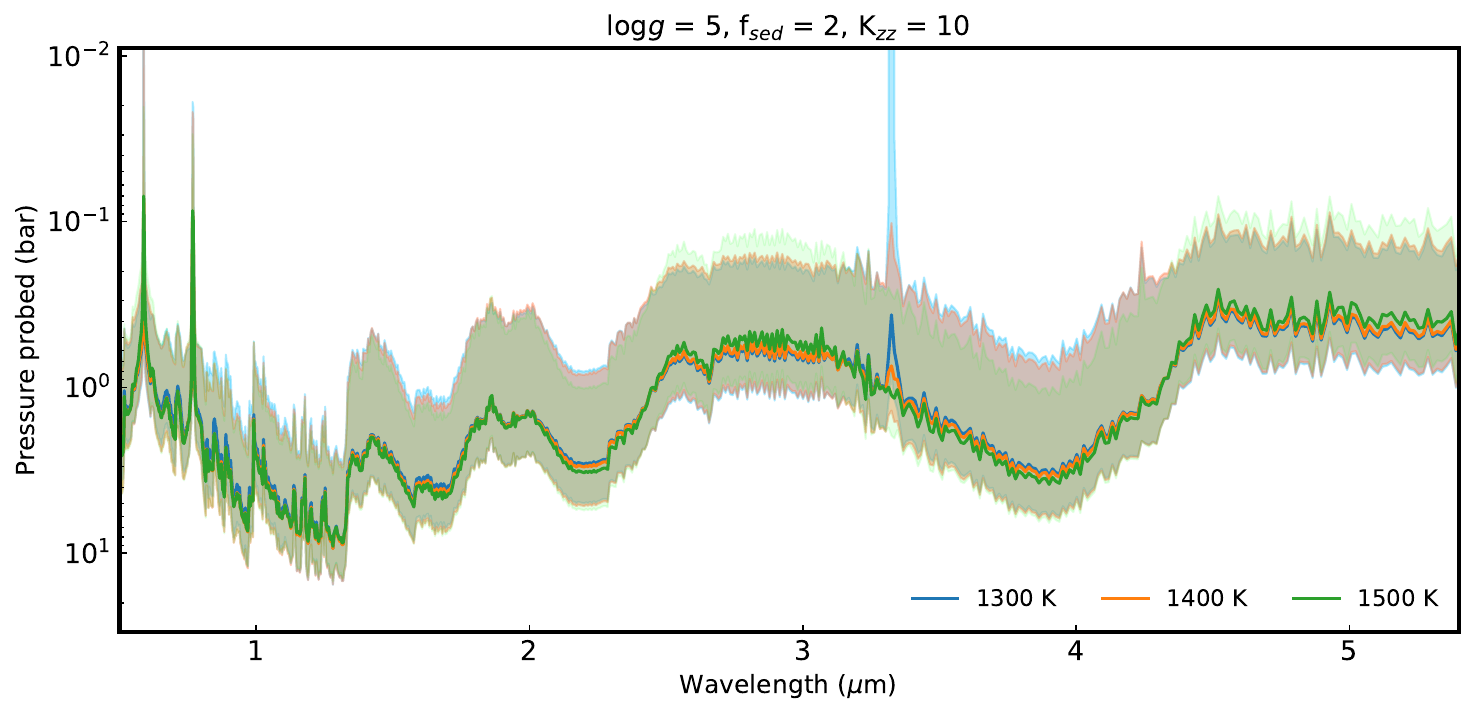}
    \caption{Contribution function using Sonora-Diamondback with parameters of T$_{eff}$=1300, 1400, 1500 K, log$g$=5, f$_{sed}=2$ and K$_{zz}$=10.}
    \label{fig:CF-difTemp}
\end{figure*}

The contribution functions for objects of temperatures ranging from 1300--1500~K and log~g of 5.0 do not significantly vary (see Figure \ref{fig:CF-difTemp}). Thus, the pressure levels traced by the wavelengths explored in this work would not change significantly, independent of the effective temperatures derived from the various atmospheric models. The only exception in which the pressure levels traced would vary is at 3.3~$\mu$m, which is not considered in the selection of molecular features for this study. We chose the contribution function corresponding to an effective temperature of 1300 K to be consistent with the results of \citealt{biller2013weather}. Also, it is significant to mention that, although our models of the contribution functions assume equilibrium chemistry, we can conclude in our interpretation of the variability/CH4 abundances found disequilibrium chemistry, mainly for the WISE 1049A object.



\begin{figure*}
    \centering
    \includegraphics[width=0.9\linewidth]{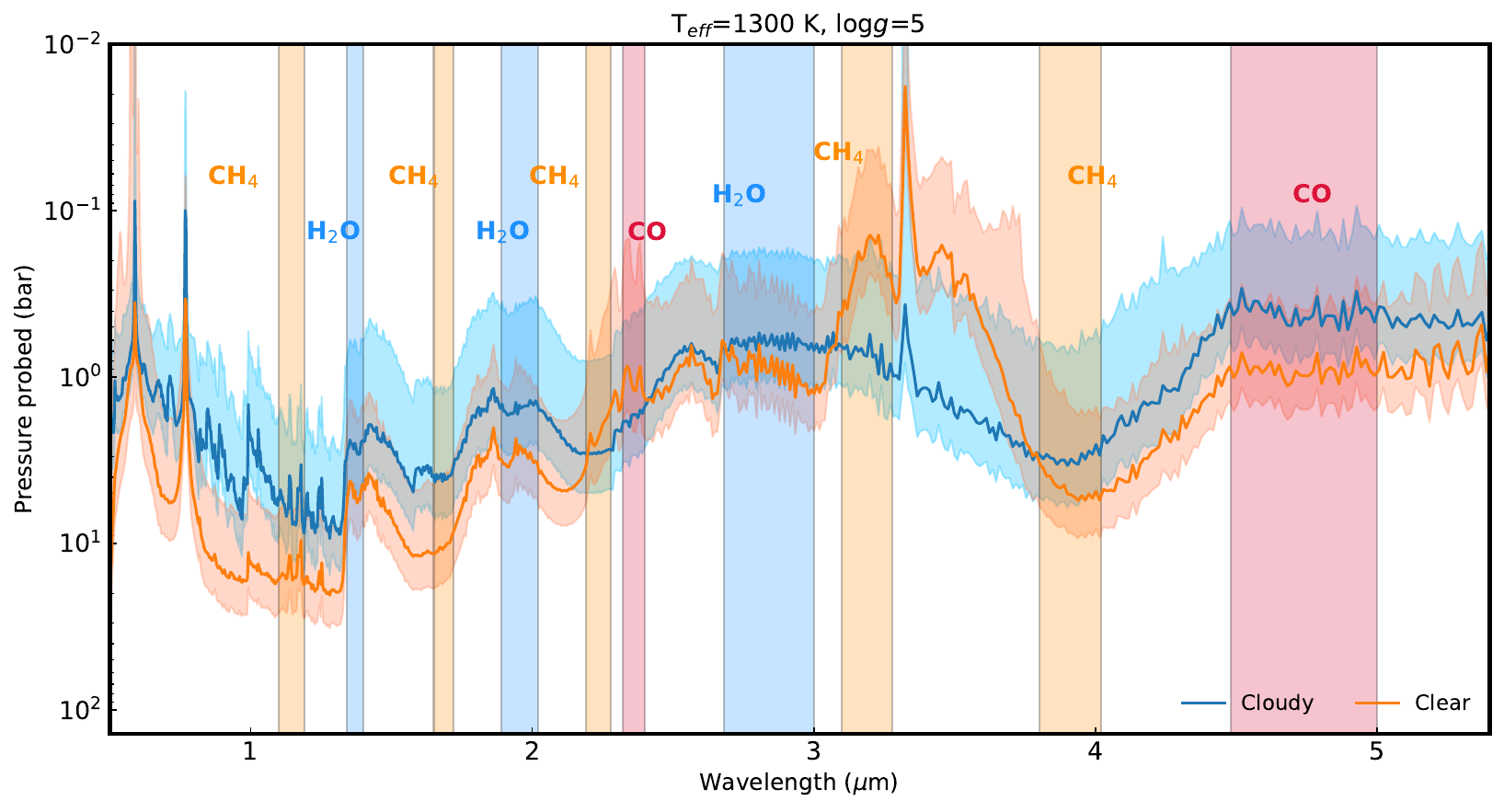}
    \caption{Contribution function using Sonora-Diamondback with parameters of T$_{eff}$=1300 K, log$g$=5, f$_{sed}=2$ and K$_{zz}$=10, for the cloudy (blue) and Sonora-Bobcat with T$_{eff}$=1300 K, log$g$=5 (orange). The marked color ranges represent the molecular bands selected for analysis in this study of H$_2$O in blue, CH$_4$ in orange, and CO in pink.}
    \label{fig:CF-cloudy-clear}
\end{figure*}

In addition, we compared the difference between the clear and cloudy atmosphere contribution functions, using Sonora-Bobcat and Sonora-Diamondback models, respectively. Figure \ref{fig:CF-cloudy-clear} shows the difference between the two contribution functions for the same parameters. In blue we see the contribution used in this analysis, considering a homogeneous cloud deck, and in orange the comparison for a clear atmosphere. The contribution function for a clear atmosphere encompasses a broader range of pressures, extending from deeper into the atmosphere (e.g., in the $J$-band) to higher altitudes (e.g., $L$-band). However, the pressure levels traced by the different molecules (shaded in colors) do not significantly change with either model. In our analysis, we continued working with the cloudy model, since it is one of the primary mechanisms that we expect to generate the variability 
(\citealt{crossfield2014global}, \citealt{karalidi2016maps}, \citealt{chen2024global}). Additionally, considering our analysis of the best fits, we are examining cloud structures in the form of planetary-scale waves that reproduce the shape of the light curves.

\end{document}